\documentclass[twocolumn,aps,superscriptaddress]{revtex4}
\usepackage{amssymb}
\usepackage{epsfig,newlfont}
\usepackage{subfigure}
\usepackage[colorlinks, 
linkcolor=blue,
urlcolor=blue,
citecolor=blue,
bookmarksopen=false]{hyperref}
\usepackage{graphicx, xcolor}
\usepackage{caption}

\newcommand{\bv}{\mathbf{v}}
\newcommand{\bE}{\mathbf{E}}
\newcommand{\bB}{\mathbf{B}}

\newcommand{\fsa}[1]{\langle{#1}\rangle}
\def\newblock{\hskip .11em plus .33em minus .07em}
\newcommand{\jlvg}[1]{\textcolor{black}{#1}}

\begin{document}

\title{Particle transport after pellet injection in the TJ-II stellarator}

\author{J.L. Velasco$^1$, K. J. McCarthy$^1$, N. Panadero$^1$, S. Satake$^2$, D. L\'opez-Bruna$^1$, A. Alonso$^1$, I. Calvo$^1$, T. Estrada$^1$, J. M. Fontdecaba$^1$, J. Hern\'andez$^1$, R. Garc\'ia$^1$, F. Medina$^1$, M. Ochando$^1$, I. Pastor$^1$, S. Perfilov$^3$, E. S\'anchez$^1$, A. Soleto$^1$, B. Ph. Van Milligen$^1$, A. Zhezhera$^4$, the TJ-II team}

\affiliation{Laboratorio Nacional de Fusi\'on, CIEMAT, Madrid, Spain}
\affiliation{National Institute for Fusion Science, Toki, Japan}
\affiliation{RNC Kurchatov Institute, Moscow, Russia}
\affiliation{Institute of Plasma Physics, NSC KIPT, Kharkov, Ukraine}

\begin{abstract}

We study radial particle transport in stellarator plasmas using cryogenic pellet injection. By means of perturbative experiments, we estimate the experimental particle flux and compare it with neoclassical simulations. Experimental evidence is obtained of the fact that core depletion in helical devices can be slowed-down even by pellets that do not reach the core region. \jlvg{This phenomenon is well captured by neoclassical predictions with DKES and FORTEC-3D.}

\end{abstract}

\maketitle 

\section{Introduction}
 
Core plasma fuelling is a critical issue for developing steady-state scenarios in fusion reactors. Gas puffing, the standard tool for creating and sustaining plasmas, will be inefficient in  large-size devices, since the particle source is located at the edge. Moreover, the particle source due to recycling  is expected to be small, since the plasma-wall interaction will be highly localized close to the divertor. Neutral beam injection (NBI)~\cite{kelley1972nbi} has also become a standard method for plasma fuelling, where the associated source can be located, at least partially, in the core region. Nevertheless, with NBI, an additional energy source is introduced into the plasma, which may be problematic from the point of view of density control. In stellarators, the reason is that particle and energy transport are coupled in the core region (see e.g. reference~\cite{beidler2011ICNTS}), as they are dominated by the neoclassical contribution~\cite{dinklage2013ncval,yokoyama2007cerc}. In particular, in Helias-type stellarators, this coupling could create hollow density profiles~\cite{maassberg1999densitycontrol} and thus exacerbate the need for fuelling mechanisms that mitigate potential core depletion. 

Pellet injection systems permit achieving relatively localized plasma fuelling without an associated energy source. For this reason, among others, they are the subject of intense research in tokamaks (see e.g.~\cite{garzotti2003jet,valovic2008mast,lang2012aug,baylor2007iter,polevoi2005iter}) and in helical devices. For instance, the pellet injection system in the Large Helical Device (LHD)~\cite{yamada2000pellets} extends its operational regime with good energy confinement~\cite{sakamoto2006repet}. More recently, a four-barrel compact pellet injector has entered operation in the stellarator TJ-II~\cite{mccarthy2008pinjector,combs2012pinjector}. In Wendelstein 7-X, in operation since December 2015~\cite{sunnpedersen2015op11}, a pellet blower gun is expected to be available during the second (OP1.2) phase of operation in order to fuel the core of plasmas heated using Electron Cyclotron Resonance Heating (ECH). As a result of this interest, comparative studies of pellet fuelling in these three devices have started~\cite{mccarthy2015ishw}. Two are the main subjects of this study: on the one hand, an effort is being made, both from the experimental and theoretical points of view, to describe the phenomena of pellet ablation. In particular, comparison of the penetration depth of the pellet with ablation models~\cite{panadero2016phd} is expected to provide predictive capability on the possibility of core fuelling in large devices. A general prediction of these models (e.g. the widely-used Neutral Gas Shielding model~\cite{parks1977ngs}, upgraded e.g. in~\cite{sakamoto2001ngs}) is that the penetration depth of the pellet may not be large enough to reach the core of dense plasmas. This increases the importance of the second line of research: transient transport of charged particles after pellet injection.

Here, perturbative experiments using small hydrogen pellets are performed on the stellarator TJ-II. Small pellets are employed for several reasons: first, it is necessary to tailor the penetration depth so that the majority of particles are deposited at an intermediate radial position. These experiments are then relatively close to reproducing the reactor-relevant situation of difficult central particle fuelling. In this situation, we will see that subsequent particle transport redistributes some of the particles to the core, compensating core depletion. This will provide experimental evidence that core depletion in helical devices can be slowed-down even by pellets that do not reach the core region. We will model this phenomenon using neoclassical theory, which describes the transport of charged particles caused by collisions combined with inhomogeneities of the magnetic field. This is an appropriate approach in plasmas in which turbulent transport is negligible (as it may be the case in the core of helical devices, see below) but also, as we will argue in Section~\ref{SEC_SCOPE}, in a more general situation: when turbulent transport is not negligible but is not affected by injection of a pellet.

Accurate comparison of neoclassical predictions with experimental measurements is the second reason for injecting small pellets. The above-mentioned prediction of potential core depletion in stellarators relies on simulations of energy and particle transport based on neoclassical theory in the core and complemented with simplified models of turbulent transport in the edge. This is supported by a previous step-by-step systematic validation of predictions of neoclassical theory with experimental results in a number of medium-sized stellarators (LHD, W7AS, TJ-II...)~\cite{yokoyama2007cerc,dinklage2013ncval}. Generally speaking, such simulations tend to achieve, with some exceptions, reasonable agreement with the experimental particle and heat fluxes within the core region. In particular, several estimates suggest that core particle transport is well described by neoclassical theory. Nevertheless, these validation activities have focused so far on electron energy~\cite{yokoyama2007cerc} and ion energy transport~\cite{dinklage2013ncval}. The reason is that particle balance studies require, at least for some devices, good knowledge of the particle source, which in turn demands a self-consistent modelling of bulk plasma, edge plasma and plasma-wall interaction, which only now is starting to become available~\cite{feng2014emc3}.

In order to overcome this problem, we will discuss perturbative experiments. Here, we describe scenarios in which modifications caused by a pellet in the plasma density and temperature profiles are minimal, and therefore the other plasma-dependent sources in the core (recycling and neutral beam injection) do not change: under these conditions, after a transient phase, the density should tend to relax back to its initial state. Measuring this density evolution with sufficient radial resolution will allow us to estimate the experimental particle fluxes associated with the perturbation without requiring knowledge of the particle sources. We will then compare them to neoclassical predictions. In particular, following recent works~\cite{satake2014eps,velasco2014eps} we will investigate the differences between the predictions of a local and monoenergetic neoclassical code, DKES~\cite{hirshman1986dkes}, and a non-local, non-monoenergetic neoclassical code, FORTEC-3D~\cite{satake2006fortec3d}, which includes higher order terms of the drift-kinetic equation. A preliminary result of previous works is that, for the plasma core, neglecting these terms may lead to overestimating the neoclassical contribution to radial ion energy transport, and therefore to underestimating the turbulent contribution to it (an effect larger for low collisionalities in devices with large ripple)~\cite{satake2015ishw}. \jlvg{In this work, we will see that the non-local neoclassical calculation is closer to the experimental measurement at outer radial positions.}

The work is organized as follows. Section~\ref{SEC_SCOPE} describes the studied scenarios, both from theoretical and operational points of view. The results are presented in Section~\ref{SEC_RES}. Finally, Section~\ref{SEC_CON} shows the conclusions.

\section{Scope of research}\label{SEC_SCOPE}

In this work, the evolution of the radial density profile after the injection of a pellet is of prime interest. Hence it is necessary to solve the flux-surface-averaged particle transport balance equation:
\begin{equation}
\frac{\partial \fsa{n}}{\partial t}+ \fsa{\nabla\cdot\Gamma} = \fsa{S}\,,\label{EQ_DNDT}
\end{equation}
where $n$ is the density, $\nabla\cdot \Gamma$ is the divergence of the radial particle flux and $S$ is the particle source. From now on, we drop the brackets $\fsa{...}$ that denote flux-surface average in order to ease the notation.

Before pellet injection ($t\!\lesssim\!t_{I}$):
\begin{equation}
\frac{\partial n}{\partial t}\bigg|_{BI}+ \nabla\cdot\Gamma_{BI} = S_{BI}\,.\label{EQ_DNDT_BI}
\end{equation}
where the subscript $BI$ denotes that all quantities are evaluated at $t\!=\!t_{BI}\!\lesssim\!t_{I}$. Then, immediately after an injection ($t\!=\!t_{AI}\!\gtrsim\!t_{I}$), the plasma density has changed and, in principle, fluxes and sources should have changed:
\begin{equation}
\frac{\partial n}{\partial t}\bigg|_{AI}+ \nabla\cdot\Gamma_{AI} = S_{AI}\,.\label{EQ_DNDT_AI}
\end{equation}
If the pellet is small enough so that the experiment can be considered perturbative, the particle source (NBI absortion, neutrals from the wall) should remain unchanged. For this, $|S_{AI}-S_{BI}|$ should be much smaller than $\left|\frac{\partial n}{\partial t}|_{AI} - \frac{\partial n}{\partial t}|_{BI}\right|$. If this is the case, then assuming that $S_{AI}\!=\!S_{BI}$ is a valid approximation, and substracting equation~(\ref{EQ_DNDT_AI}) and equation~(\ref{EQ_DNDT_BI}), yields:
\begin{equation}
  \frac{\partial n}{\partial t}\bigg|_{AI} - \frac{\partial n}{\partial t}\bigg|_{BI} = \nabla\cdot(\Gamma_{BI}-\Gamma_{AI})\,.\label{EQ_DNDT_PERT}
\end{equation}

The left-hand-side of equation~(\ref{EQ_DNDT_PERT}) can be determined from the experiment  while the right-hand side can be calculated using neoclassical theory. The goal of this work is to measure, calculate and compare particle fluxes in the transient phase after the injection of a pellet. Good agreement is expected if particle transport is dominated by the neoclassical contribution, but note that this is a sufficient but not necessary condition: as long as the turbulent contribution to $\nabla\cdot(\Gamma_{BI}-\Gamma_{AI})$ is small enough (i.e., as long as turbulent transport is small or, if it is large, has not changed), neoclassical transport determines the transient core density evolution.

Finally, pellet injection also has the effect of slightly reducing the electron and ion temperatures of the radial positions where ablation occurs, thus allowing similar perturbative studies to be done for energy transport of both species. Such studies are left for future work.

\subsection{Measurement of transient evolution: experimental set-up}\label{SEC_EXP}
 
TJ-II is a four-period, low magnetic shear stellarator device with an average minor radius $a$ of approximately 0.2$\,$m, a major radius of 1.5$\,$m, plasma volume about 1$\,$m$^3$, and magnetic field on axis $B(0)\!\approx\!1\,$T~\cite{sanchez2013tj-ii}. For this work, plasmas created with hydrogen in the standard magnetic configuration are heated using two gyrotrons operated at 53.2$\,$GHz, the 2nd harmonic of the electron cyclotron resonance frequency, with $P_{ECH} \leq 500\,$kW. Central electron densities $n_e(0)$ and electron temperatures $T_e(0)$ up to $1.7\times 10^{19}\,$m$^{-3}$ and $1\,$keV, respectively, are achieved. Plasmas last up to 250$\,$ms. In addition, plasmas have been created and maintained using a neutral beam injector heater that provides up to 520$\,$kW. As a result, plasmas with $n_e(0)\le 3\times 10^{19}\,$m$^{–3}$ and $T_e(0)\le 380\,$eV were achieved. With a lithium coating on the vacuum vessel wall, these plasmas last up to 120$\,$ms.

The pellet formation, acceleration, guide line diagnostic, delivery and control systems were developed, built and tested at the laboratories of the Fusion Energy Division of Oak Ridge National Laboratory, Tennessee, USA~\cite{combs2012pinjector}. For these experiments, pellets containing between $5\times 10^{18}$ and $10^{19}$ hydrogen atoms are injected at speeds between 800 and 1200$\,$m/s, their mass being determined with an in-line microwave cavity diagnostic and their ablation being followed using silicon diodes with Balmer H$_\alpha$ transmission filters ($\lambda_0 = 656\,$nm): these diodes view the pellet through the plasma from above and behind the flight path and allow penetration depth to be obtained. The variability of pellet mass and velocity are below 10\% and 3\% respectively; since penetration is determined principally by plasma temperature and, next, by pellet velocity (see e.g.~\cite{parks1977ngs}), the present conditions ensure reproducible plasma penetration, see~\cite{mccarthy2015epsd,mccarthy2015cwgm}.

 In addition, TJ-II is equipped with numerous diagnostics. A Thomson Scattering (TS) system~\cite{herranz2003TS} provides one set of radial density and temperature profiles per discharge; hence shot-to-shot reconstruction is used to determine their evolution. The line-averaged electron density, $\overline{n_e}$, is obtained with 10~$\mu$s temporal resolution using a microwave interferometer~\cite{sanchez2004interferometer}. Since the interferometer cannot capture the fast rise time in the line-averaged electron density after an injection, it is necessary to calibrate it using TS data.

Moreover, core $T_e$ measurements ($r/a\!<\!0.4$) are made along a discharge using a soft X-ray filtered diagnostic (SXR)~\cite{baiao2010SXR} (this system provides estimates of $T_e$ slightly below the TS measurement, but within the error bars of the diagnostics~\cite{baiao2012SXR}). Finally, the temporal evolution of the core ion temperature $T_i$ is obtained using a Neutral Particle Analyzer (NPA)~\cite{fontdecaba2014NPA}. Although not shown in this paper, H$\alpha$ monitors and a Baratron-type manometer are used for estimating the particle source.

\subsection{Calculation of the flux: codes}\label{SEC_NC}

In this subsection, the two codes used in the calculation of the neoclassical contribution to the particle flux are briefly presented.

FORTEC-3D~\cite{satake2006fortec3d} is a $\delta f$ radially global Monte Carlo code that solves the drift kinetic equation (DKE) for both electrons and ions including finite-orbit-width effects in three-dimensional configurations. For each species, we have:

\begin{eqnarray}
(\mathbf{v_\parallel}+{{\mathbf{v_{M}}}}+{{\mathbf{v_E}}})\cdot\nabla \delta f+{{\frac{\mathrm{d}K}{\mathrm{d}t}\frac{\partial \delta f}{\partial K}}}-{{C(\delta f)}} =\nonumber\\ -\left(\mathbf{v_M}\cdot\nabla +\frac{\mathrm{d}K}{\mathrm{d}t}\frac{\partial}{\partial K}\right) f_M+{{\cal{P}}f_M}\,,\label{EQ_DKE}
\end{eqnarray}
where $\delta f(r,\zeta,\theta,K,\mu)$ is the deviation of the distribution function with respect to the Maxwellian, $r$ is the radial coordinate, $\zeta$ and $\theta$ are the toroidal and poloidal Boozer angles, $K$ is the kinetic energy normalized to that of the thermal ions and $\mu$ is the magnetic moment. In the first term of the left-hand-side, $v_\parallel$ is the parallel velocity and
\begin{eqnarray}
\mathbf{v_M}&=&\frac{2K-\mu B}{Ze}\frac{\mathbf{B}\times\nabla B}{B^3}\,,\nonumber\\
\mathbf{v_E}&=&\frac{\mathbf{E}\times\mathbf{B}}{B^2}\,,
\end{eqnarray}
are the magnetic drift and the $\bE\times \bB$ drift caused by the radial electric field $E_r$ respectively, and $Ze$ the particle charge. $C(\delta f)+\cal{P}(f_M)$ is a linearized Fokker-Planck collision operator with pitch-angle as well as energy scattering and momentum conservation (more details can be found in reference~\cite{satake2006fortec3d}). The particle flux can be calculated by taking moments of the distribution function and flux-surface-averaging:
\begin{equation}
\Gamma=\bigg\langle\int\mathrm{d}\bv \delta f \bv_M\cdot\nabla r\bigg\rangle\,,
\end{equation}
The radial electric field is set by ambipolarity of the neoclassical fluxes, $\Gamma_e(E_r)=\Gamma_i(E_r)$ for a pure hydrogen plasma ($n_e\!=\!n_i$). Note that only here we use subindexes to make explicit the species-dependence.

DKES~\cite{hirshman1986dkes} solves the DKE in the local and monoenergetic limit: starting from equation~(\ref{EQ_DKE}), one neglects the magnetic drift and the variation of the kinetic energy on the left-hand-side of the equation, and takes the so-called incompressible limit of the $\bE\times \bB$ drift, $\mathbf{v_E}=\frac{\mathbf{E}\times\mathbf{B}}{\fsa{B^2}}$. Neglecting the magnetic drift is generally a good approximation, since the $\bE\times\bB$ drift is larger in an inverse-aspect-ratio expansion; nevertheless, when $E_r$ is close to zero, the poloidal component of the magnetic drift becomes important. \jlvg{This effect is more important the closer the radial electric field is to zero, the lower the collisionality and the larger the magnetic ripple, i.e., the larger the $1/\nu$ flux~\cite{beidler2011ICNTS}}. Finally, the collision operator is simplified to the Lorentz operator (pitch-angle-scattering only), so the kinetic energy becomes a parameter rather than a variable. The particle flux can then be calculated as
\begin{equation}
\Gamma = -n D_1\left[\left(\frac{n'}{n}-\frac{Ze E_r}{T}\right)+\left(\frac{D_2}{D_1}-\frac{3}{2}\right)\frac{T'}{T}\right]\,.\label{EQ_GAMMANC}
\end{equation}
The neoclassical thermal transport coefficients $D_1$ and $D_2$, can be calculated by convolution of solutions of the monoenergetic drift-kinetic equation calculated with DKES. More details on this calculation for the specific case of the TJ-II magnetic configuration are found in reference~\cite{velasco2011bootstrap}. 

Of the two presented numerical codes, DKES will be used for comparison between neoclassical predictions and experimental measurements, since it is much less time consuming; the accuracy of DKES for these plasmas will be checked by benchmarking with FORTEC-3D for one of the cases. Note that the flux can be written as in equation~(\ref{EQ_GAMMANC}), as a linear combination of the gradients of density, temperature and electrostatic potential, provided that the radial magnetic drift term of equation~(\ref{EQ_DKE}) is small. This is generally true even in cases where $E_r$ is close to zero and the poloidal magnetic drift matters. Therefore we can still discuss neoclassical transport in terms of plasma gradients even if the approximations included in DKES do not hold. Obviously, if we were to calculate the coefficients $D_1$ and $D_2$ with FORTEC-3D, they would differ from those calculated with DKES.

For some cases included in this study, the particle source is estimated. There are three contributions: gas puffing, particle deposition due to NBI and recycling due to the wall. The second one is estimated using the code FAFNER~\cite{teubel1994fafner}; the third one, using the neutral transport code EIRENE~\cite{reiter2001eirene}. The inputs for EIRENE are the plasma profiles, the quantity of gas coming from puffing and NBI systems and a particle confinement time $\tau$ that reflects the level of flux to the wall, i.e., $S=S(n,T_e,T_i,$ gas input$,\tau)$. An additional equation for $\tau$ is thus needed, which can be written as:
\begin{equation}
\tau(n,S)=\frac{\int_Vn\mathrm{d}V}{\int_\cal{S}\Gamma\mathrm{d}\cal{S}}\,,\label{EQ_TAU}
\end{equation}
where $\cal{S}$ and $V$ are the area and volume of the last-closed flux-surface. Note that, once the source has been calculated, equation~(\ref{EQ_DNDT}) can be used for estimating the experimental flux, if $\frac{\partial n}{\partial t}$ can be measured accurately or if it is negligible compared with the other two terms in the equation.

\subsection{Plasma scenarios}\label{SEC_SCE}

The properties of the two scenarios studied in this work are summarized in this subsection. Firstly, scenario I is a situation with core depletion and difficult central fuelling. The target plasma is heated by NBI, has a relatively peaked density profile and the line-averaged density decreases at a constant rate at the time of pellet injection due to a reduction of the gas-puffing. Looking at equation~(\ref{EQ_DNDT_PERT}), we will be in a situation in which $\frac{\partial n}{\partial t}|_{BI}\!<\!0$. A small pellet is injected, so that it ablates at an intermediate radial position. Then the fuelled plasmas have flatter core density, hence $\nabla\cdot(\Gamma_{BI}-\Gamma_{AI})>0$ due to the change in the thermodynamical force associated with the density gradient (see equation~(\ref{EQ_GAMMANC})). For a large enough change, there will be indirect core fuelling, indicated experimentally by $\frac{\partial n}{\partial t}|_{AI}>0$.

Next, scenario II is an ECH-heated plasma with a hollow density profile. Density is kept approximately constant in time, i.e. $\frac{\partial n}{\partial t}|_{BI}=0$. The pellet then ablates close to the core, the hollowness is removed, we end up with a peaked density profile; hence $\nabla\cdot(\Gamma_{BI}-\Gamma_{AI})<0$, and there will be a fast density evolution with $\frac{\partial n}{\partial t}|_{AI}<0$ back to the initial situation. One has to note that, in this low density plasma, the number of injected particles is not negligible when compared to that in the target plasma, so the assumption of perturbative experiment may be not good enough for a quantitative comparison.

\section{Results}\label{SEC_RES}

\subsection{Scenario I}

\begin{figure}
\begin{center}
\includegraphics[angle=0,width=\columnwidth]{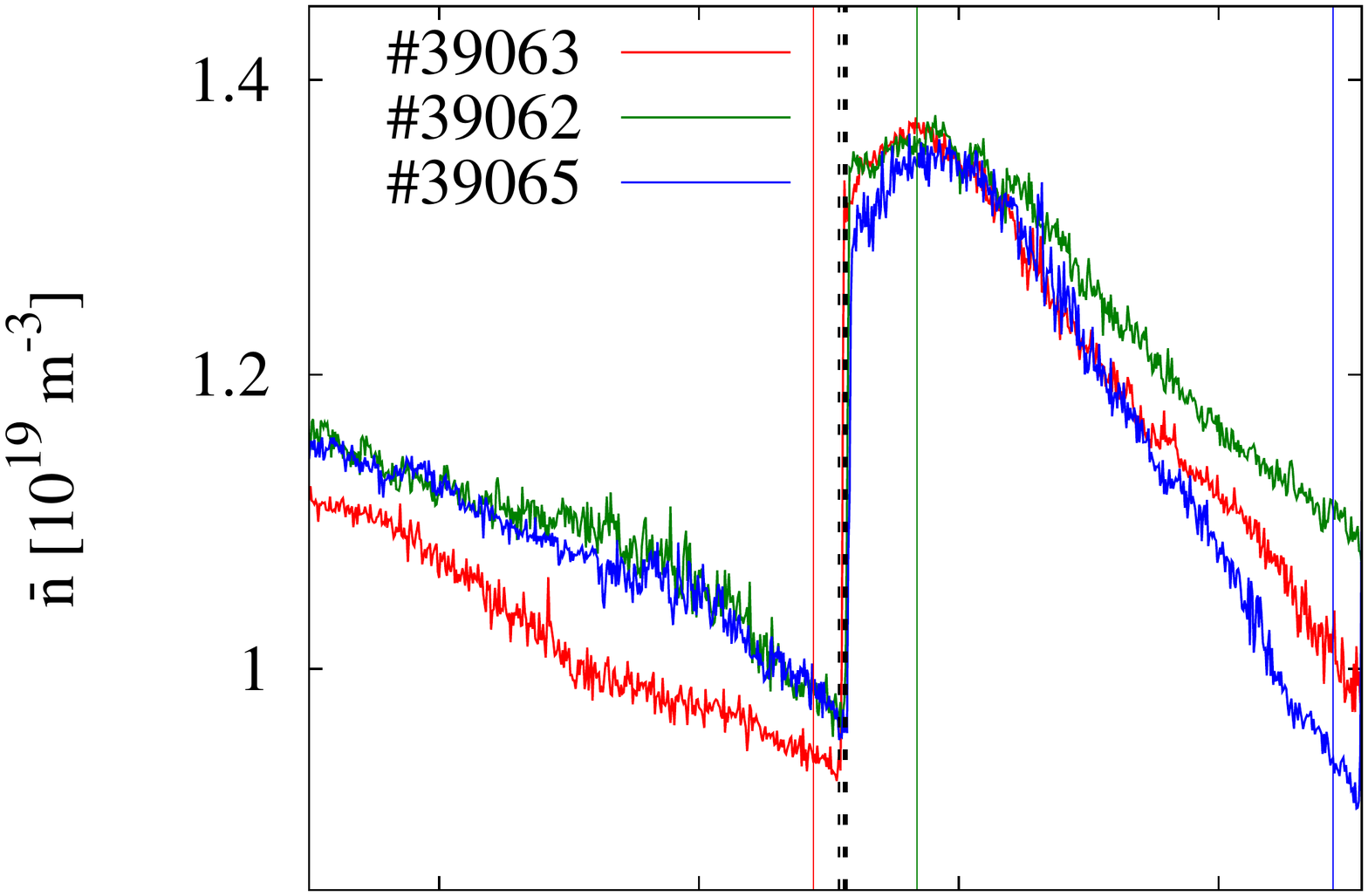}\vskip-2.1cm
\includegraphics[angle=0,width=\columnwidth]{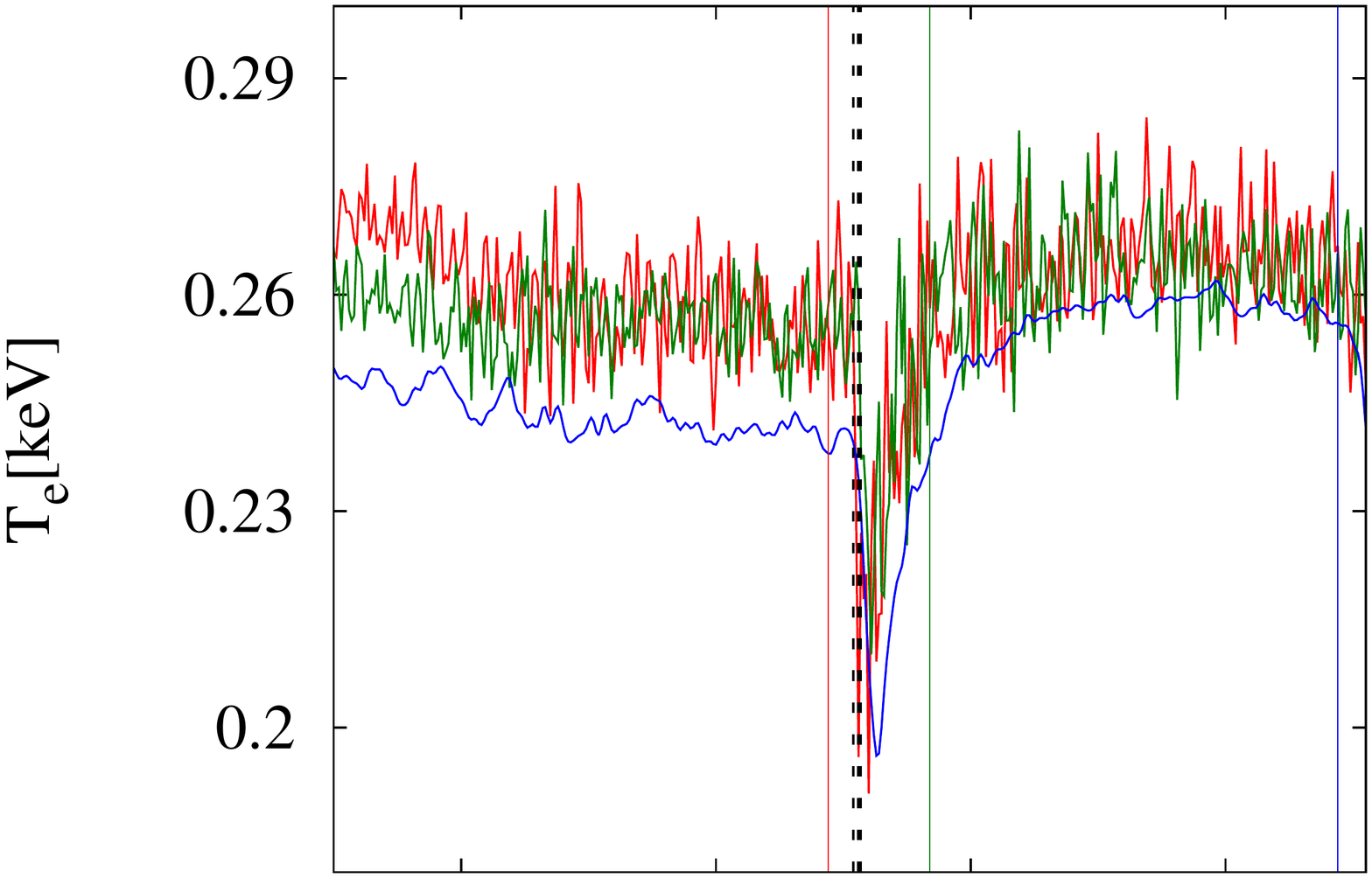}\vskip-2.1cm
\includegraphics[angle=0,width=\columnwidth]{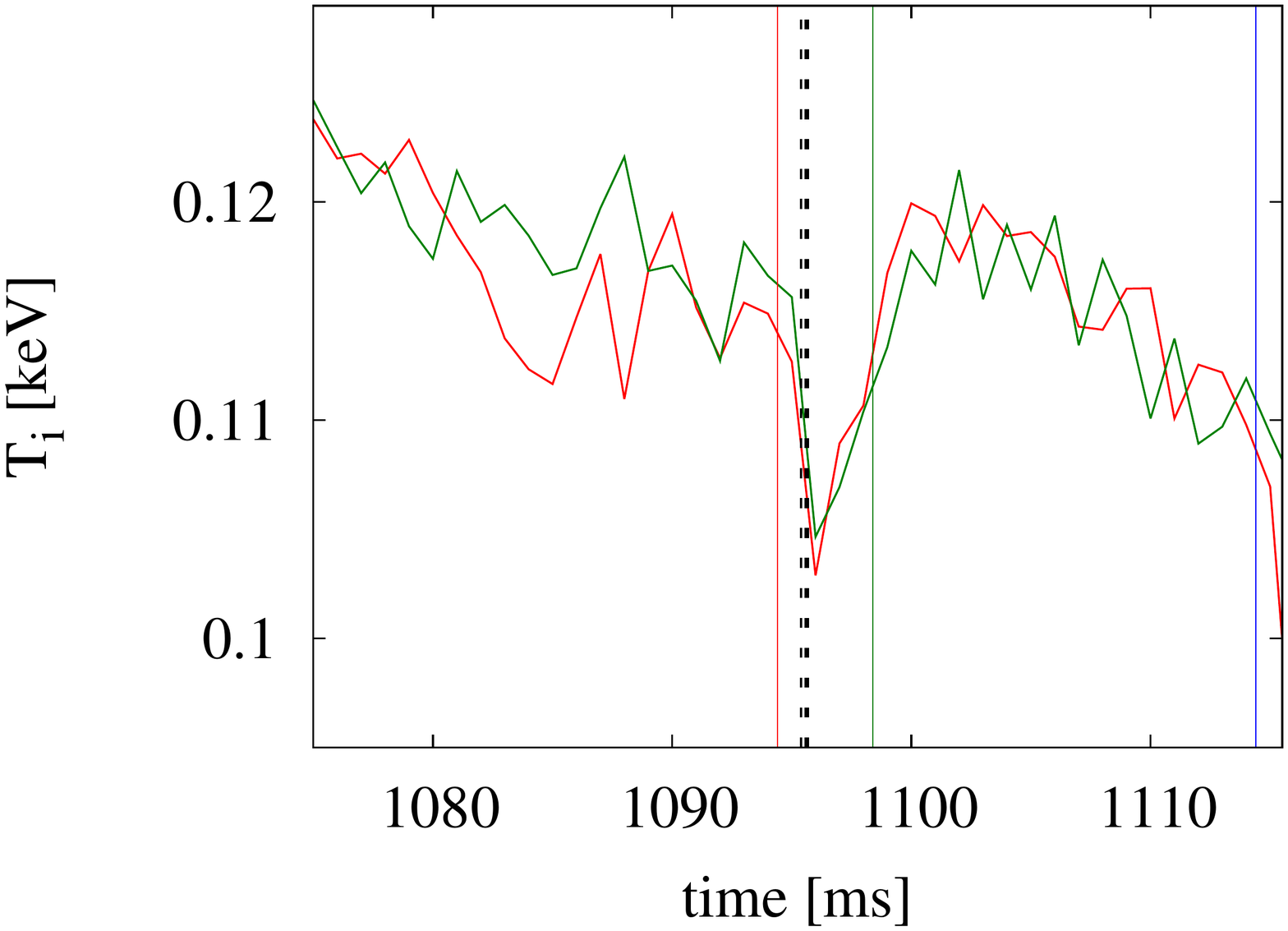}
\end{center}
\vskip-1cm\caption{Time traces of (top) line-averaged density, (center) central electron temperature, and (bottom) central ion temperature in discharges \#39063 (red), \#39062 (green) and \#39065 (blue). Pellet ablation is indicated by the sudden increase of density and a black vertical line; coloured vertical lines show the TS time for each discharge.}
\label{FIG_TRACES1}
\end{figure}

Figure~\ref{FIG_TRACES1} shows the temporal evolution of the main plasma parameters along discharges \#39062, \#39063 and \#39065: line-average density, and central electron and ion temperatures. Before injection of the pellet, the temperatures are approximately constant with time, and the density decreases at an approximately constant rate $\frac{\partial n}{\partial t}|_{BI}\approx - 10^{20}\,$m$^{-3}$s$^{-1}$. Pellets containing $6.5\times 10^{18}\!\pm\!2\%$ hydrogen atoms are injected at $1020\!\pm\!0.5\%\,$m/s. The ablation of the pellet is observed at $t\!=\!1095\,$ms for the three discharges. Both central temperatures show a $\sim 25\,$\% decrease and return to the previous level after $5\,$ms. On the other hand, the evolution of the density is much slower, denoting that the particle confinement time is larger than the energy confinement time for this scenario. Hence, this (rough) scale separation allow us to focus on particle transport while ignoring energy transport. 

\begin{figure}
\begin{center}
\includegraphics[angle=0,width=\columnwidth]{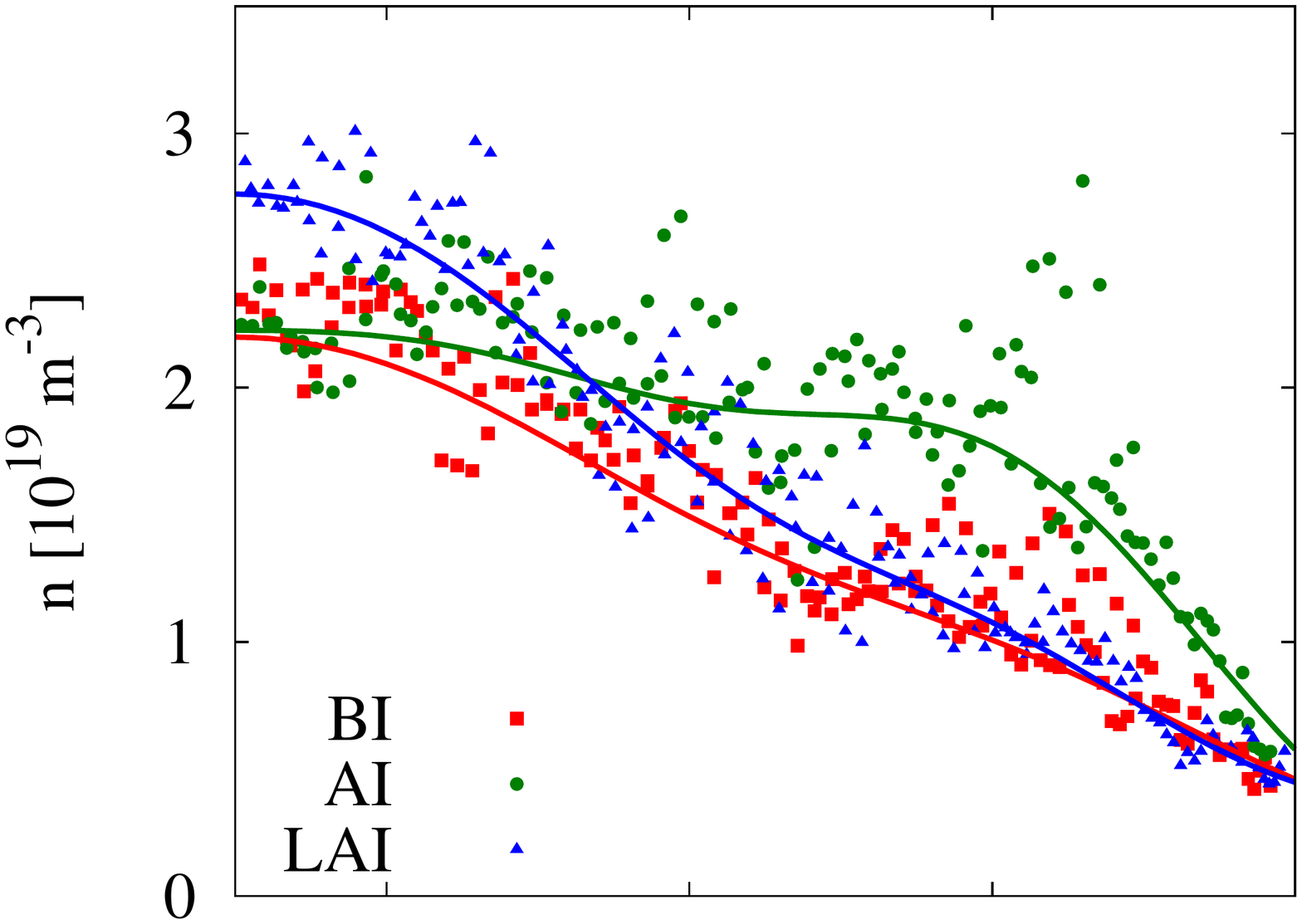}\vskip-2.1cm
\includegraphics[angle=0,width=\columnwidth]{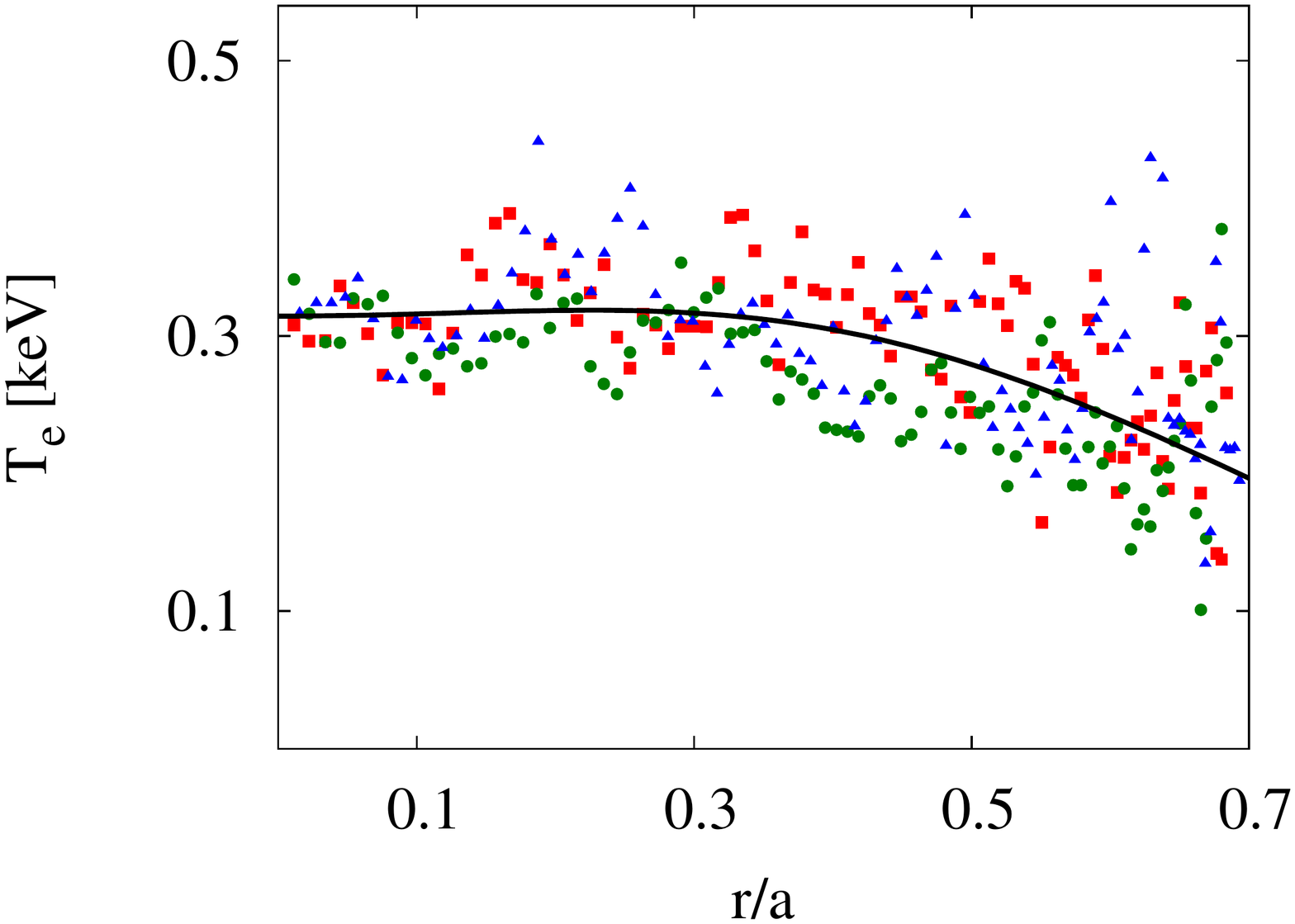}\vskip-2.1cm
\end{center}
\vskip-1.cm\caption{Electron density and temperature profiles before pellet injection (BI, \#39063), immediately after pellet injection (AI, \#39062), and long after pellet injection (LAI, \#39065).}
\label{FIG_TS1}
\end{figure}

The above three discharges are used to reconstruct the evolution of the profiles (similar sets of three discharges can be found in~\cite{mccarthy2015epsd,mccarthy2015cwgm}). Figure~\ref{FIG_TS1} shows the density and electron temperature profiles immediately before (\#39063), immediately after (\#39062) and 20$\,$ms after (\#39065) injection of a pellet. Most of the pellet ablation takes place about $r\!\approx\!0.5a$, as indicated by the change of density profile in figure~\ref{FIG_TS1} (top) and by the local $H_\alpha$ monitors. This profile has a distinctive temporal evolution: after an initial increase in density about the ablation radii, the density increases in the core region ($r\!<\!0.3a$), while it decreases in outer regions ($0.3a\!<\!r\!<\!0.7a$). Since the core density remains almost unchanged by the injection, as does the energy source, the core electron temperature returns to previous values, and it does so very quickly, as anticipated by figure~\ref{FIG_TRACES1}. Altogether, these measurements indicate that the  situation is one of difficult central fuelling, and that the problem is partially mitigated by subsequent particle transport.

\begin{figure}
\begin{center}
\includegraphics[angle=0,width=\columnwidth]{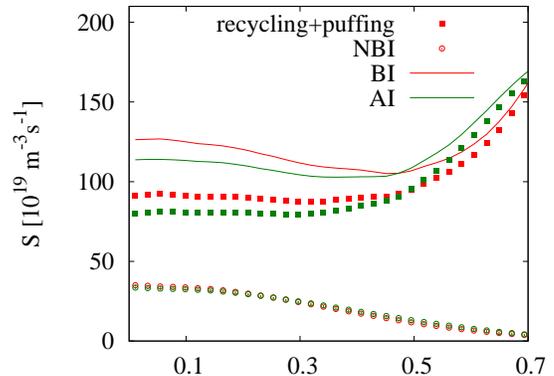}
\end{center}
\vskip-1.cm\caption{Particle sources for discharges before (BI, \#39063) and after (AI, \#39062) pellet ablation.}
\label{FIG_SOURCES1}
\end{figure}

Before discussing theoretically this experimental result, it is necessary to confirm that the experiment can be considered to be perturbative. For this, the particle sources $S$ are estimated before injection and immediately after pellet complete ablation (so that there is no source associated to it anymore). This is seen in figure~\ref{FIG_SOURCES1}, where the contributions of the NBI system and wall recycling are plotted. The difference in the core $|S_{AI}-S_{BI}|$ is between 0 and 10$^{20}\,$m$^{-3}$s$^{-1}$. This difference, due to changes in recycling, is smaller than $\left|\frac{\partial n}{\partial t}|_{AI}-\frac{\partial n}{\partial t}|_{BI}\right|$: inspection of figure~\ref{FIG_TS1} shows that the core density increases by $\sim 0.5\times 10^{19}\,$m$^{-3}$ in 16$\,$ms, i.e. $\left|\frac{\partial n}{\partial t}|_{AI}\right|\approx 3\times 10^{20}\,$m$^{-3}$s$^{-1}$ and $\left|\frac{\partial n}{\partial t}|_{AI}-\frac{\partial n}{\partial t}|_{BI}\right|\approx 4\times 10^{20}\,$m$^{-3}$s$^{-1}$.

\begin{figure}
\begin{center}
\includegraphics[angle=0,width=\columnwidth]{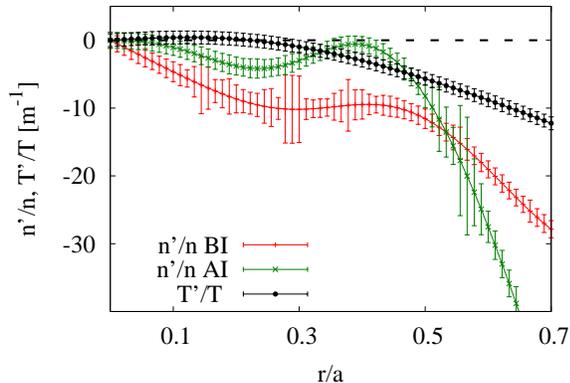}
\end{center}
\vskip-1.cm\caption{Thermodynamical forces associated to the density and temperature gradients before and after the injection in scenario I.}
\label{FIG_TF1}
\end{figure}

We thus proceed with the neoclassical simulations. As input for the calculations we used smoothed profiles taken from figure~\ref{FIG_TS1}. Given the above-mentioned different time-scales, the electron and ion temperatures are approximately constant in time and equal to their values before pellet injection during much of the density evolution that we want to model. We thus neglect the small differences and provide, for the sake of simplicity a common fit for $T_e(r)$; $T_i(r)$, only measured in the core region, is taken proportional to $T_e(r)$~\cite{fontdecaba2010pfr}. As discussed in Section~\ref{SEC_SCE}, after injection, the density profile becomes flatter at the core ($r<0.5a$) and steeper at outer radial positions ($0.5a\!<\!r\!<\!0.7a$). In order to emphasize that the change in the density gradient is significant, in figure~\ref{FIG_TF1} we show the thermodynamical forces associated with the density and temperature gradients before and after injection, as calculated from Thomson Scattering measurements using a Bayesian method~\cite{milligen2011bayes}. The density gradient is the main drive for particle flux, both before and after injection of a pellet, which supports our previous approximation of considering constant $T_e(r)$ during the transient evolution. The error bars are smaller than the change caused by the pellet injection.

\begin{figure}
\begin{center}
\includegraphics[angle=0,width=\columnwidth]{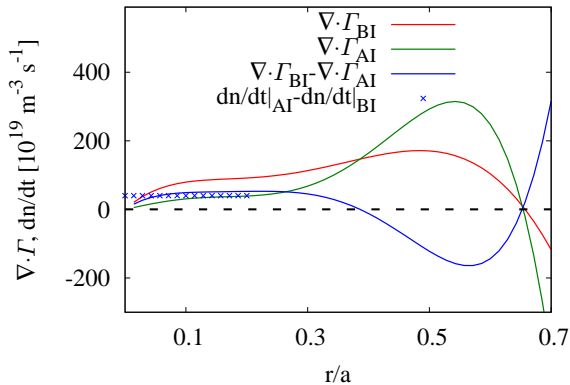}
\end{center}
\vskip-1.cm\caption{Contribution of neoclassical transport, calculated with DKES, to the particle balance equation in scenario I.}
\label{FIG_DNDT1}
\end{figure}

The consequences are seen in figure~\ref{FIG_DNDT1}, where we show the divergence of the neoclassical flux calculated before and after the injection of the pellet, and the predicted density evolution according to equation~(\ref{EQ_DNDT_PERT}). We focus on the region $r/a\!<\!0.6$, where neoclassical theory generally describes successfully radial transport in helical devices~\cite{dinklage2013ncval}. As one could expect by inspection of equation~(\ref{EQ_GAMMANC}), it is positive in the core region ($r\!<\!0.4a$) and negative in outer positions, as measured in the experiment. The value of the core density increase is also in rough agreement with measurements, since we predict $\nabla\cdot(\Gamma_{BI}-\Gamma_{AI})$ between 0 and $5\times 10^{20}\,$m$^{-3}$s$^{-1}$ in the core region, compared with $\frac{\partial n}{\partial t}|_{BI}-\frac{\partial n}{\partial t}|_{AI}$ between 0 and $4\times 10^{20}\,$m$^{-3}$s$^{-1}$ measured by Thomson Scattering, and represented by crosses in Fig.~\ref{FIG_DNDT1}.

\begin{figure}
\begin{center}
\includegraphics[angle=0,width=\columnwidth]{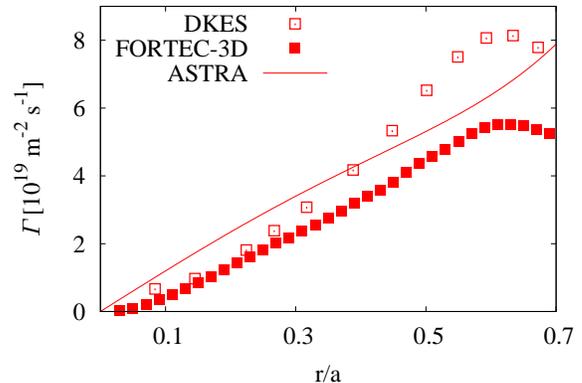}
\end{center}
\vskip-1.cm\caption{Radial fluxes of discharge \#39063.}
\label{FIG_GAMMA1}
\end{figure}

\begin{figure}
\begin{center}
\includegraphics[angle=0,width=\columnwidth]{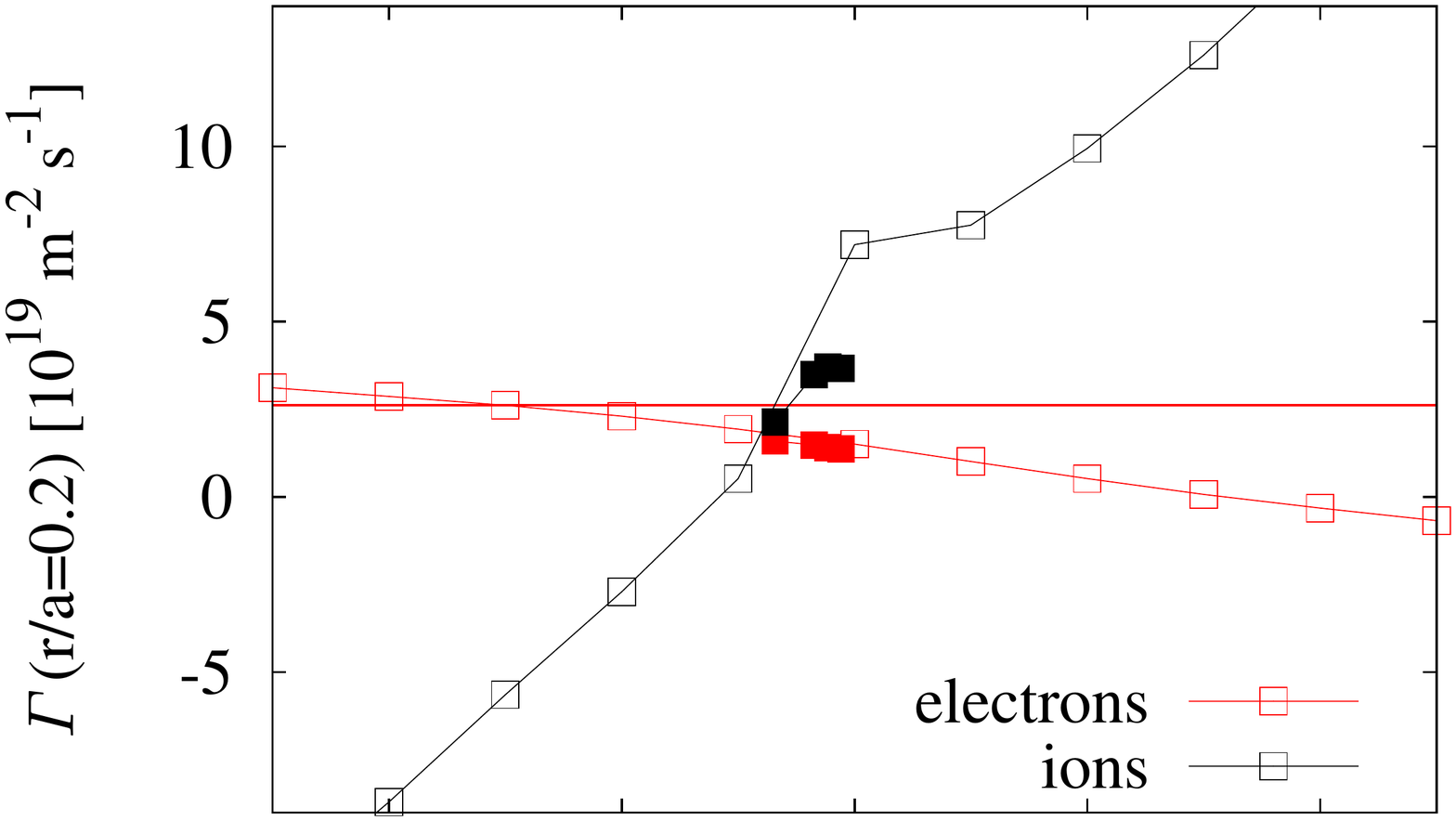}\vskip-2.5cm
\includegraphics[angle=0,width=\columnwidth]{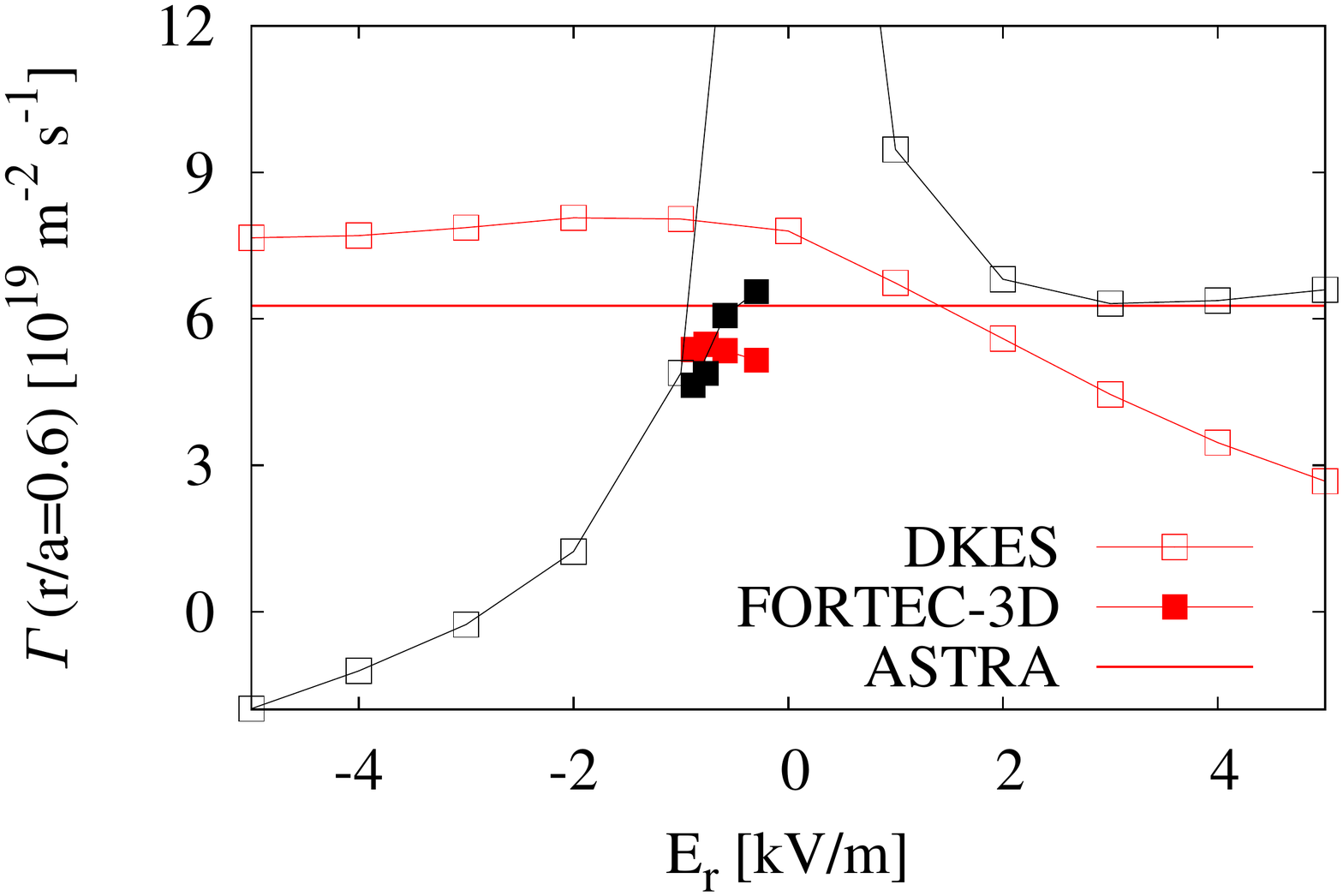}
\end{center}
\vskip-1.3cm\caption{\jlvg{Neoclassical particle radial fluxes as a function of the radial electric field for discharge \#39063 at r/a=0.2 (top) and r/a=0.6 (bottom). The crossing between the electron and ion curves gives the neoclassical estimate of the particle flux, which can then be compared to the experimental estimate. Red corresponds to electrons and black to ions; open squares to DKES and filled squares to FORTEC-3D; a line with the flux calculated with ASTRA is shown for reference.}}
\label{FIG_AMB1}
\end{figure}

It should be noted that the quantitative agreement reported above allows us to conclude that density evolution {caused by} injection of a pellet is dominated by neoclassical transport, {which was the main subject of study of this work (and using equation (4) has allowed us to discuss it without having to rely on estimates of the particle source)}; however, it provides no information of whether it was also the case before the injection. In order to check that {for the sake of completeness}, we have to {estimate the source and} compare directy the experimental fluxes with the neoclassical predictions. Figure~\ref{FIG_GAMMA1} shows the experimental flux for discharge \#39063, before the pellet injection. It is calculated with the transport suite ASTRA~\cite{pereverzev2002ASTRA}, using equation~(\ref{EQ_DNDT}) and the source from figure~\ref{FIG_SOURCES1}. \jlvg{We compare it with the neoclassical fluxes calculated with DKES and FORTEC-3D and achieve a reasonable agreement in the core region $r/a\!<\!0.4$ (the error bars in the thermodynamical forces of Fig.~\ref{FIG_TF1} provide a rough estimate of the precision of the neoclassical calculations; the $\sim$ 30\% uncertainty in the calibration of the H$_\alpha$ monitor, used for the evaluation of the particle source, provides an estimate of the precision of the calculation with ASTRA). At outer radial positions, $0.4\!<\!r/a\!<\!0.7$, the radial flux predicted with DKES is above the experimental estimate, which would suggest that DKES overestimates neoclassical transport under these plasma conditions (since otherwise, if the DKES calculation were not an overestimate of neoclassical transport, we would need an additional inward anomalous pinch for the total flux to be in agreement with the experimental estimate). Indeed FORTEC-3D predicts a radial flux about 50\% smaller than the one calculated with DKES, and below the experimental estimate. The reason behind the difference between the two simulations is sketched in Fig.~\ref{FIG_AMB1}: the inclusion of the magnetic drift term in the drift kinetic equation removes the peak close to $E_r\!=\!0$ in the radial flux of both species. At $r/a\!=\!0.6$ (bottom) the peak is relatively large, leading to relevant differences between DKES and FORTEC-3D; at $r/a\!=\!0.2$ (top), since the collisionality is larger and the ripple smaller, these differences are smaller.}

A more comprehensive comparison between theory and experiment for this discharge can be found in reference~\cite{satake2015ishw,satake2016ppcf}: there, poor agreement was found between predictions of $E_r$ and measurements of electrostatic potential by means of HIBP. Note nevertheless that, in these NBI ion-root plasmas, the particle flux is basically determined by that of the electrons, which depends weakly on $E_r$~\cite{satake2015ishw,satake2016ppcf}. Therefore, while this disagreement would cast doubts on predictions of energy transport for discharge \#39063, it is not the case for particle transport.  Finally, the fact that neoclassical predictions of particle flux agree with two independent estimates of the experimental flux gives confidence for the three of them.

\subsection{Scenario II}

\begin{figure}
\begin{center}
\includegraphics[angle=0,width=\columnwidth]{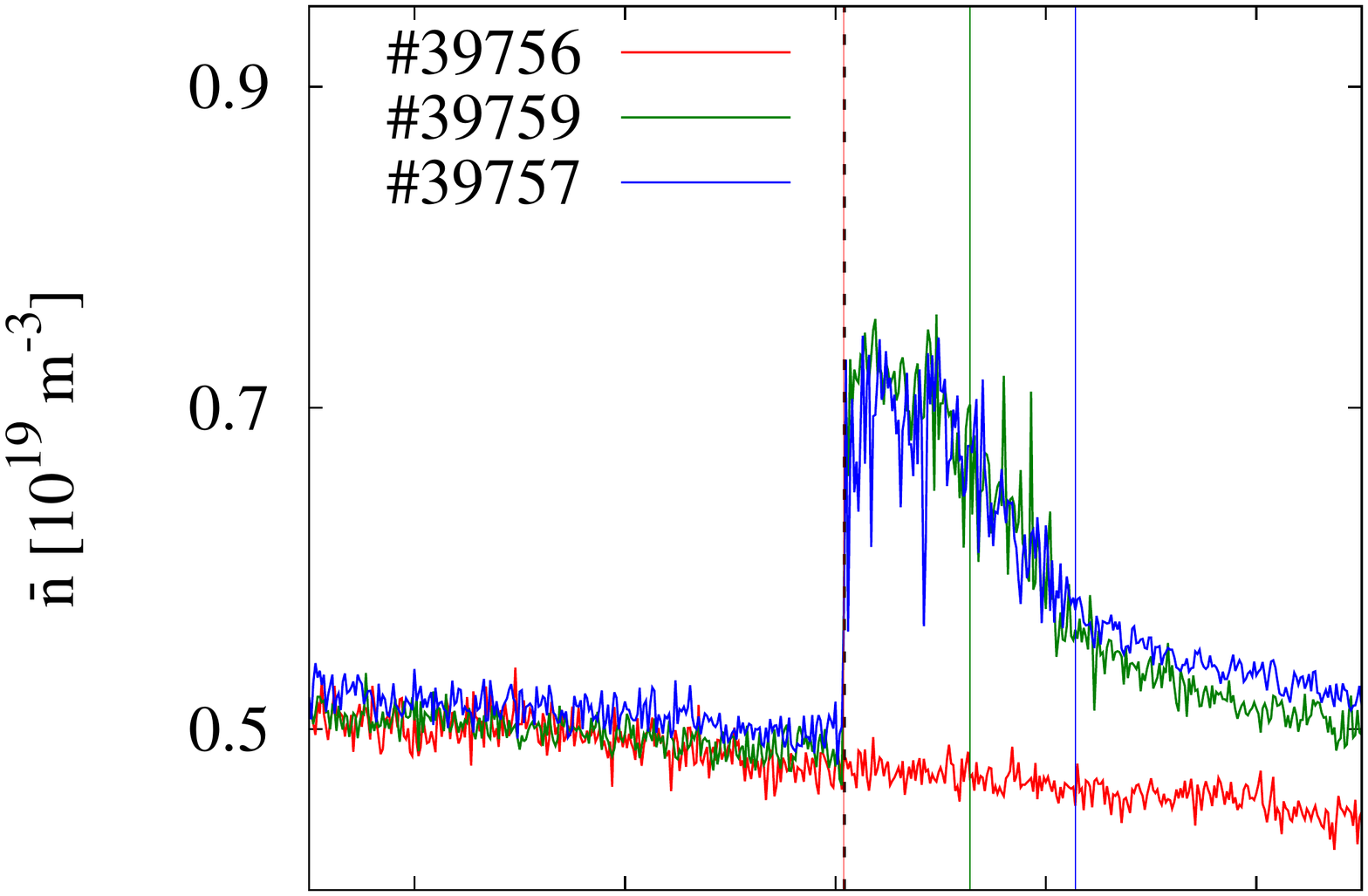}\vskip-2.1cm
\includegraphics[angle=0,width=\columnwidth]{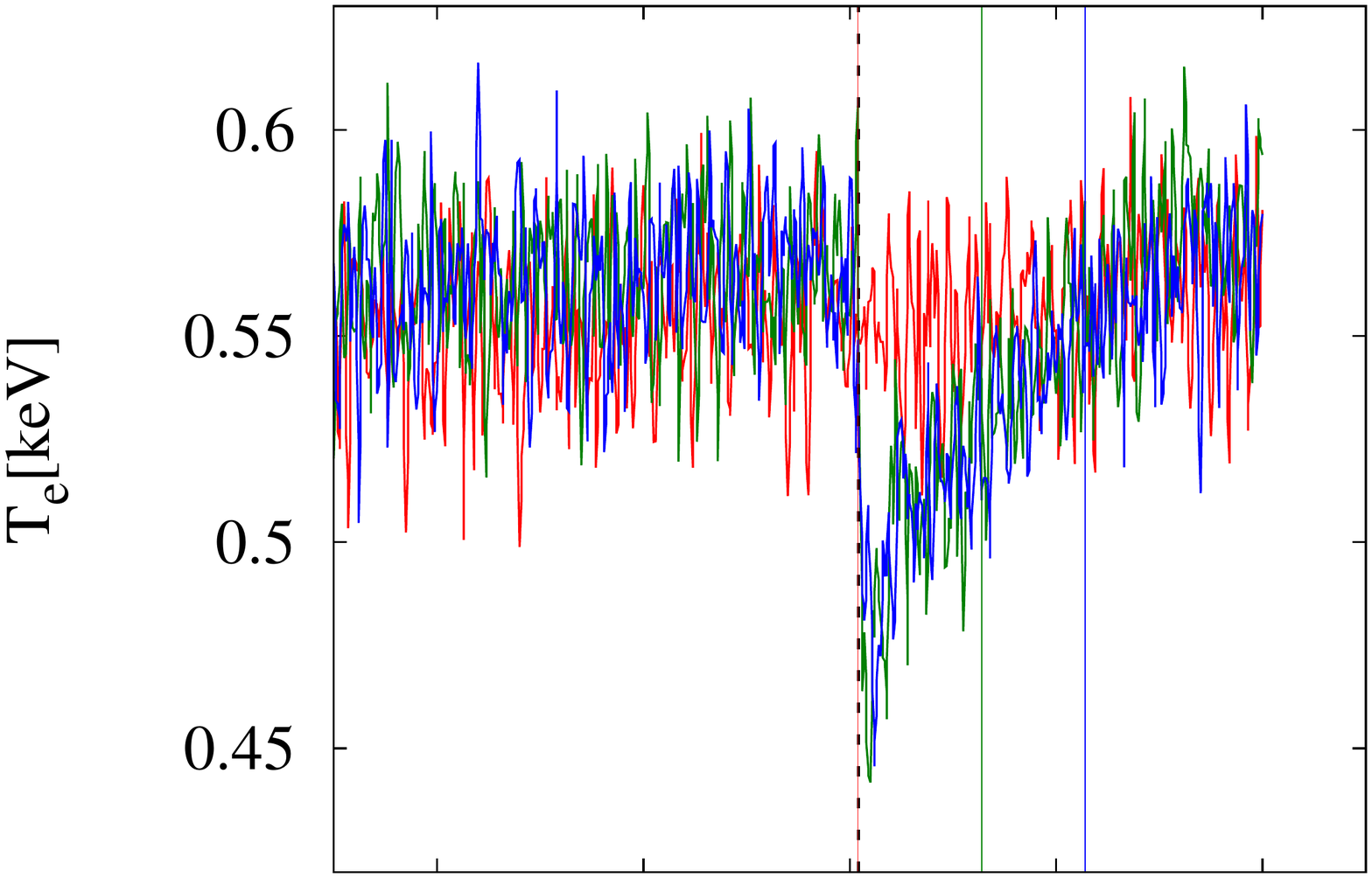}\vskip-2.1cm
\includegraphics[angle=0,width=\columnwidth]{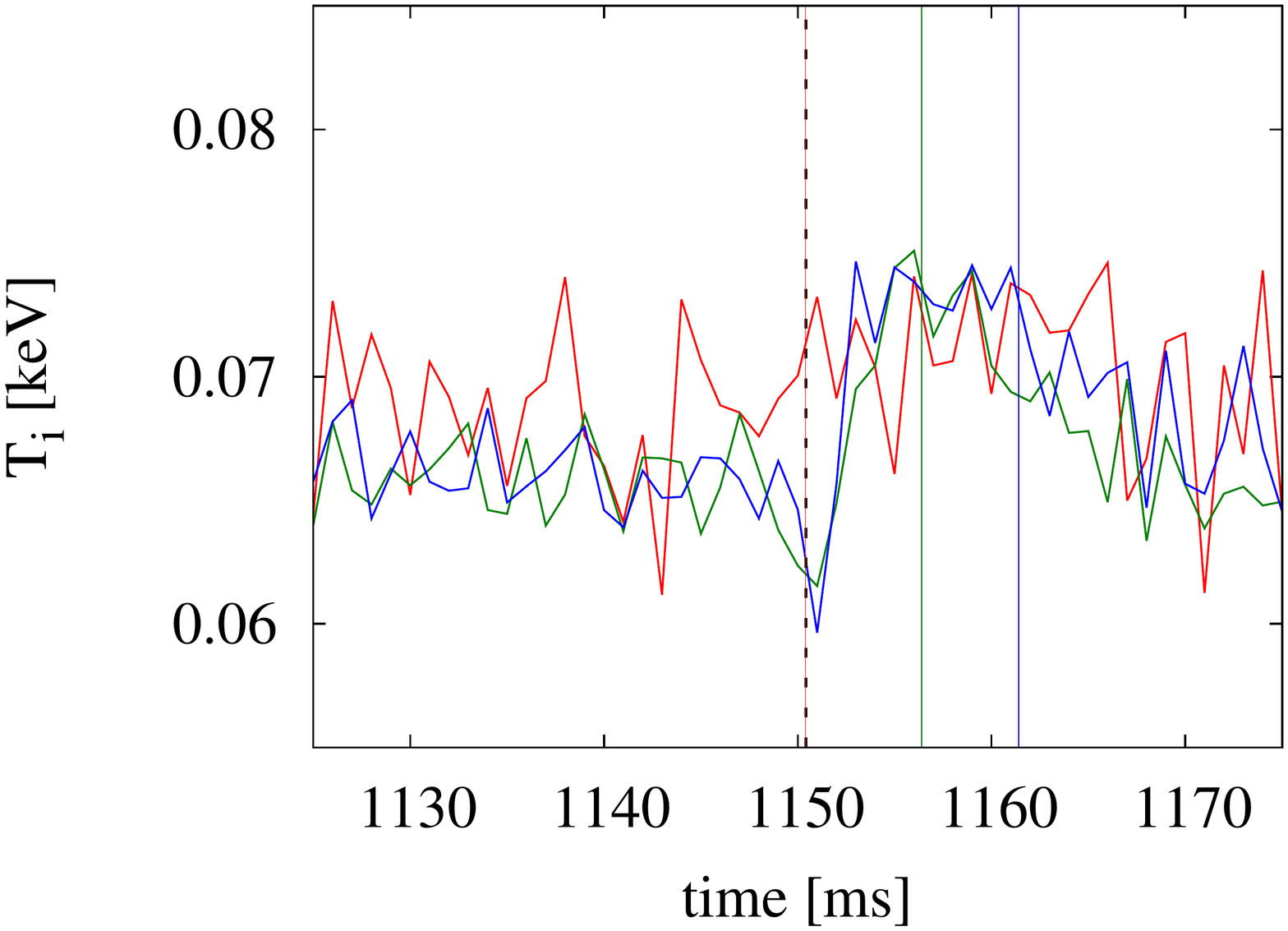}
\end{center}
\vskip-1cm\caption{Time traces of (top) line-averaged density, (center) central electron temperature, and (bottom) central ion temperature in discharges \#39756 (red), \#39759 (green) and \#39757 (blue). Pellet ablation is indicated by the sudden increase of density and a black vertical line; coloured vertical lines show the TS time for each discharge.}
\label{FIG_TRACES2}
\end{figure}

Figure~\ref{FIG_TRACES2} is the equivalent of figure~\ref{FIG_TRACES1} for scenario II: it shows the temporal evolution of the line-averaged density, and central electron and ion temperatures along discharges \#39756, \#39759 and  \#39757. The situation is comparable to that of the NBI plasma: before the injection of the pellet, the temperatures are approximately constant with time, and the density decreases at a very small constant rate, $\frac{\partial n}{\partial t}|_{BI}\approx -10^{19}\,$m$^{-3}$s$^{-1}$. Pellets containing $7.3\times 10^{18}\!\pm\!2\%$ hydrogen atoms are injected at $950\!\pm\!0.6\%\,$m/s. The ablation of the pellet takes place at $t\!=\!1150\,$ms for the latter two. The electron temperature decreases by about 20\%, while the ion temperature increases by about 20\%. The latter may be partially caused by the larger thermal coupling to the electrons, consequence of the increase of $n$, or a small change in ion energy transport. Both temperatures return to the values before injection in some 15$\,$ms. As in the NBI case, the evolution of the density is slower, although not much slower.

\begin{figure}
\begin{center}
\includegraphics[angle=0,width=\columnwidth]{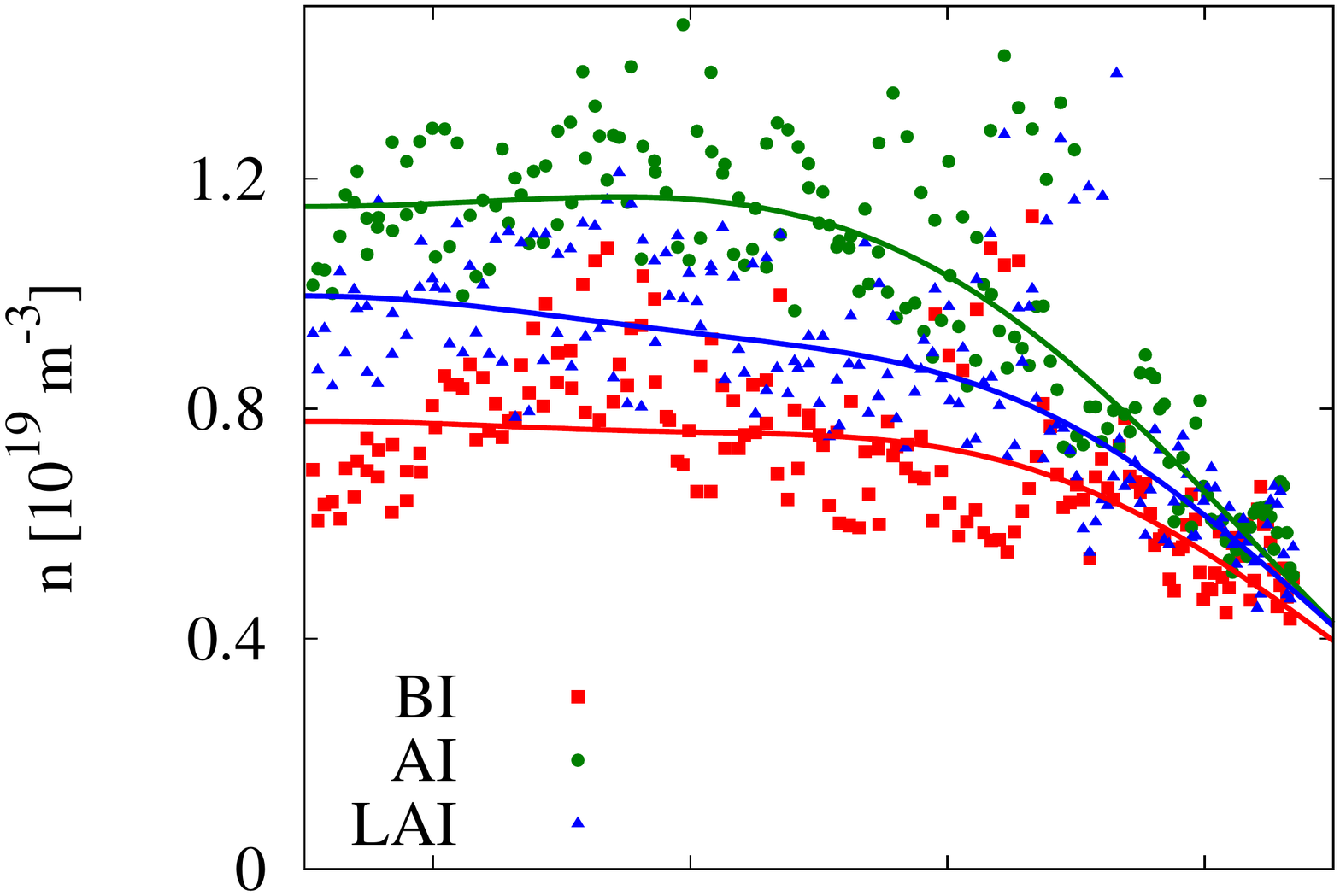}\vskip-2.1cm
\includegraphics[angle=0,width=\columnwidth]{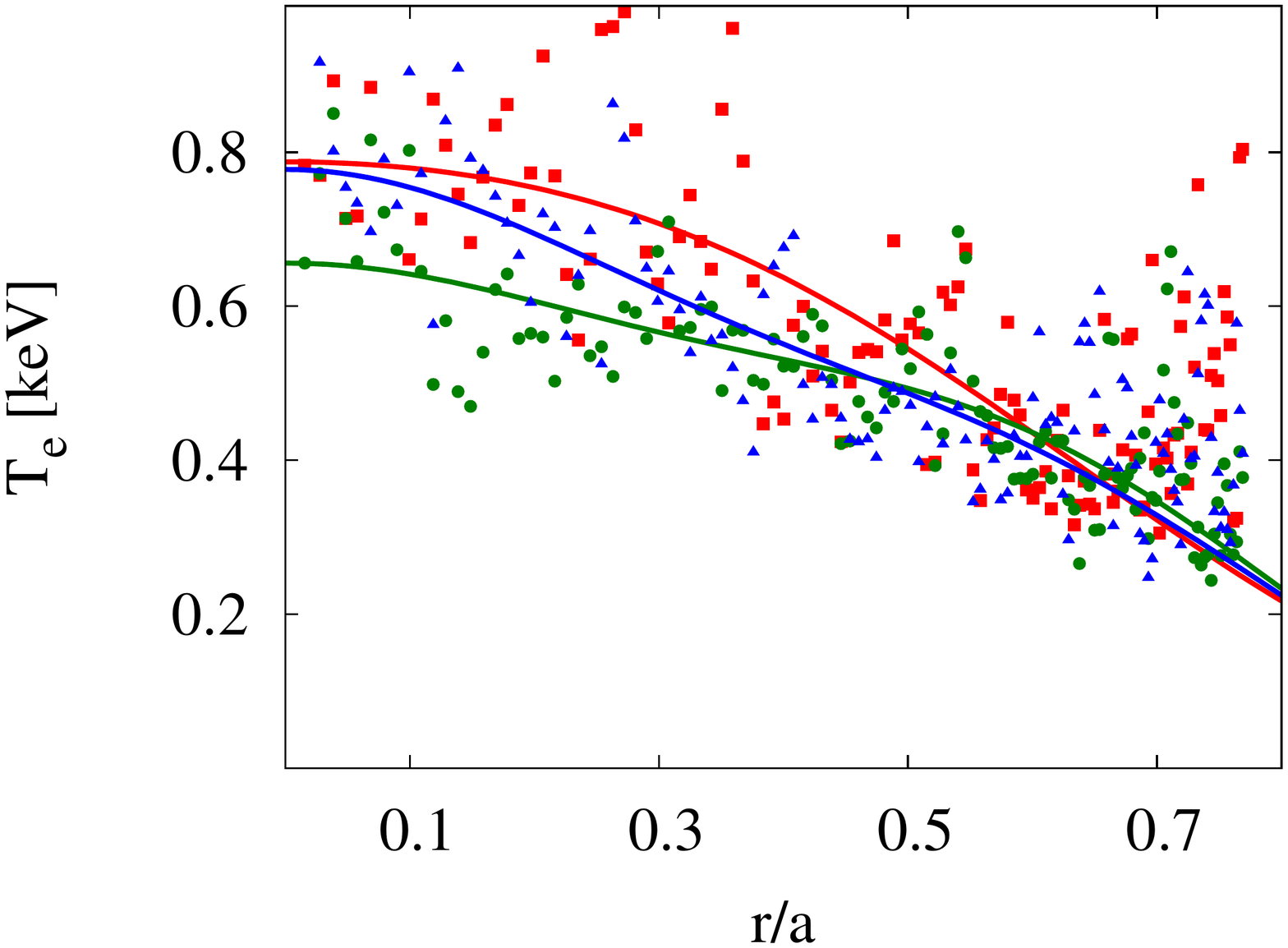}\vskip-2.1cm
\end{center}
\vskip-1.cm\caption{Electron density and temperature profiles before pellet injection (BI, \#39756), immediately after pellet injection (AI, \#39759), and long after pellet injection (LAI, \#39757).}
\label{FIG_TS2}
\end{figure}

The above three discharges are used to reconstruct the evolution of the profiles, as we did for scenario I. Figure~\ref{FIG_TS2}, comparable to figure~\ref{FIG_TS1}, shows the density and electron temperature profiles immediately before (\#39756), 6$\,$ms after (\#39759) and 11$\,$ms after (\#39757) injection of a pellet (actually discharge \#39756 has no pellet injection). The pellet is ablated at $r\!<\!0.8a$, as indicated by the increase in $n_e$. As shown in figure~\ref{FIG_TRACES2}, the electron temperature evolution is slower than in the NBI case, and differences between the $T_e$ profiles are not negligible: the electron temperature is reduced by the pellet at $r\!<\!0.8a$ and then returns faster to previous values in the core, where most of the ECH power is deposited. The density profile behaves in an opposite manner to that of scenario I. After an initial incresase at $r\!<\!0.8a$, the density decreases monotonously in all the core region. Note that no change in the profile shape is created by the pellet: it was slightly hollow before injection, and it remains so after it (the hollowness of the density profile very close to the magnetic axis is not considered significative). 

In these low-density plasmas, the change in density is large, and this in turn may modify the electron energy source (as $P_{ECH}/\overline{n}$ changes), so it is difficult to justify that the experiment is perturbative. Moreover, these ECH plasmas are close to a \textit{critical} density at which a transport bifurcation takes place at TJ-II, caused by a change of root of the radial electric field (see e.g. reference~\cite{velasco2012prl} and refereces therein). Nevertheless, the time evolution of discharges \#39757 and \#39759 is similar to that of a perturbative experiment: the line-averaged density and the electron temperature show an approximately exponential relaxation back to their initial values.

\begin{figure}
\begin{center}
\includegraphics[angle=0,width=\columnwidth]{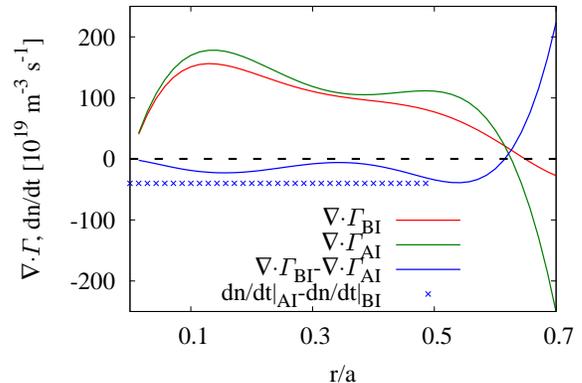}
\end{center}
\vskip-1.cm\caption{Transient density evolution calculated with DKES and estimated with Thomson Scattering in scenario II.}
\label{FIG_DNDT2}
\end{figure}

With this in mind, we repeat the simulations performed for scenario I, again focusing on $r/a\!<\!0.7$. In this case, differences in the $T_e$ profiles before and after injection must be taken into account in the simulations; the ion temperature profile is assumed proportional to the density profile~\cite{fontdecaba2010pfr}. Since the density profile shape has not fundamentally changed, the main difference between the density before and after ablation is the absolute value, not the gradient. After ablation, density is larger, so is ion collisionality and, for plasmas in electron root with ions in the $\sqrt{\nu}$ regime, the particle flux. This is shown in figure~\ref{FIG_DNDT2}: the divergence of the neoclassical flux calculated before and after the injection of the pellet, and the predicted density evolution according to equation~(\ref{EQ_DNDT_PERT}) are plotted. We predict a density decrease faster than the initial rate, and in reasonable quantitative agreement with the experiment, also represented in figure~\ref{FIG_DNDT2}: we obtain $\nabla\cdot(\Gamma_{BI}-\Gamma_{AI})$ between 0 and $-3\times 10^{20}\,$m$^{-3}$s$^{-1}$ in the core region, while TS yields $\frac{\partial n}{\partial t}|_{AI}-\frac{\partial n}{\partial t}|_{BI}\!=\!-4\times 10^{20}\,$m$^{-3}$s$^{-1}$.

\section{Conclusions}\label{SEC_CON}

In this paper, it is demonstrated that neoclassical theory can account for the particle evolution after the injection of a cryogenic pellet. \jlvg{In particular, improved neoclassical simulations show better agreement with experiment at outer radial positions}. This is relevant due to the foreseen problem of density control in large helical devices, and the predicted difficulty of core fuelling. 

In TJ-II plasmas, in which a particle source exists ($S\!\ne\!0$) and the density variation is negative but small in absolute value ($\frac{\partial n}{\partial t}|_{BI}\!\lesssim\!0$)  the core density can be made to transiently increase ($\frac{\partial n}{\partial t}|_{AI}\!>\!0$) by injection of a pellet at an intermediate radial position. Note that this is independent of whether the gradient density is positive or not. 

Since this result can be understood on the grounds of neoclassical theory, one may try to generalize it: in a stellarator reactor, in which the particle source is expected to be negligible ($S\!=\!0$) and one could have core depletion ($\frac{\partial n}{\partial t}|_{BI}\!<\!0$), the injection of a pellet could mitigate the density loss $\frac{\partial n}{\partial t}|_{AI}> \frac{\partial n}{\partial t}|_{BI}$. Nevertheless, complete density control ($\frac{\partial n}{\partial t}|_{AI}\!\ge\!0$) could require positive density gradients, as measured e.g. in LHD~\cite{mccarthy2015ishw}. In order to provide more support to these conclusions, further studies of particle transport should be performed in studies closer to reactor-relevant conditions, e.g., in W7-X plasmas.

As a conclusion, we have obtained experimental evidence of the fact that, in some scenarios, core depletion can be compensated or slowed-down even by pellets that do not reach the core region, a phenomenon that is well captured by neoclassical theory. Future work will include study of energy transport and of particle transport in NBI plasmas in which the pellet produces positive density gradients.

\section*{Acknowledgments}

The authors are indebted to Dr E. de la Luna for valuable discussions on pellet studies in tokamaks, and to Dr. R. Sakamoto for spotting a mistake in section 3. This work has been carried out within the framework of the EUROfusion Consortium and has received funding from the Euratom research and training programme 2014-2018 under grant agreement No 633053. The views and opinions expressed herein do not necessarily reflect those of the European Commission. This research was supported in part by grants ENE2012-30832 and ENE2013-48679, Ministerio de Econom\'ia y Competitividad, Spain. FORTEC-3D simulations were carried out with HELIOS supercomputer system at International Fusion Energy Research Centre, Japan, under the Broader Approach collaboration between Euratom and Japan, and with Plasma Simulator in NIFS, under the NIFS Collaboration Research program (NIFS15KNST079). All the discharges presented in this work are available at the International Stellarator-Heliotron Profile Database (ISHPDB), \url{https://ishpdb.ipp-hgw.mpg.de/}~\url{http://ishpdb.nifs.ac.jp/index.html}.

%\bibliographystyle{unsrt}
%\bibliography{/Users/velasco/Work/PAPERS/bibliography}

\begin{thebibliography}{10}

\bibitem{kelley1972nbi}
G~G Kelley, O~B Morgan, L~D Stewart, W~L Stirling, and H~K Forsen.
\newblock {Neutral-beam-injection heating of toroidal plasmas for fusion
  research}.
\newblock {\em Nuclear Fusion}, 12(2):169, 1972.

\bibitem{beidler2011ICNTS}
C~D Beidler, K~Allmaier, M~Yu Isaev, S~V Kasilov, W~Kernbichler, G~O Leitold,
  H~Maa{\ss}berg, D~R Mikkelsen, S~Murakami, M~Schmidt, D~A Spong, V~Tribaldos,
  and A~Wakasa.
\newblock Benchmarking of the mono-energetic transport coefficients. results
  from the international collaboration on neoclassical transport in
  stellarators (icnts).
\newblock {\em Nuclear Fusion}, 51(7):076001, 2011.

\bibitem{dinklage2013ncval}
A~Dinklage, M~Yokoyama, K~Tanaka, J~L Velasco, D~L\'opez-Bruna, C~D Beidler,
  S~Satake, E~Ascas\'ibar, J~Ar\'evalo, J~Baldzuhn, Y~Feng, D~Gates, J~Geiger,
  K~Ida, M~Jakubowski, A~L\'opez-Fraguas, H~Maassberg, J~Miyazawa, T~Morisaki,
  S~Murakami, N~Pablant, S~Kobayashi, R~Seki, C~Suzuki, Y~Suzuki, Yu~Turkin,
  A~Wakasa, R~Wolf, H~Yamada, M~Yoshinuma, {LHD Exp. Group}, {TJ-II Team}, and
  {W7-AS Team}.
\newblock {Inter-machine validation study of neoclassical transport modelling
  in medium- to high-density stellarator-heliotron plasmas}.
\newblock {\em Nuclear Fusion}, 53(6):063022, 2013.

\bibitem{yokoyama2007cerc}
M~Yokoyama, H~Maa{\ss}berg, C~D Beidler, V~Tribaldos, K~Ida, T~Estrada,
  F~Castej\'on, A~Fujisawa, T~Minami, T~Shimozuma, Y~Takeiri, A~Dinklage,
  S~Murakami, and H~Yamada.
\newblock {Core electron-root confinement (CERC) in helical plasmas}.
\newblock {\em Nuclear Fusion}, 47(9):1213, 2007.

\bibitem{maassberg1999densitycontrol}
H~Maa{\ss}berg, C~D Beidler, and E~E Simmet.
\newblock {Density control problems in large stellarators with neoclassical
  transport}.
\newblock {\em Plasma Physics and Controlled Fusion}, 41(9):1135, 1999.

\bibitem{garzotti2003jet}
L~Garzotti, X~Garbet, P~Mantica, V~Parail, M~Valovič, G~Corrigan, D~Heading,
  T~T~C Jones, P~Lang, H~Nordman, B~P\'egouri\'e, G~Saibene, J~Spence,
  P~Strand, J~Weiland, and {contributors to the EFDA-JET Workprogramme}.
\newblock {Particle transport and density profile analysis of different JET
  plasmas}.
\newblock {\em Nuclear Fusion}, 43(12):1829, 2003.

\bibitem{valovic2008mast}
M~Valovič, K~Axon, L~Garzotti, S~Saarelma, A~Thyagaraja, R~Akers, C~Gurl,
  A~Kirk, B~Lloyd, G~P Maddison, A~W Morris, A~Patel, S~Shibaev, R~Scannell,
  D~Taylor, M~Walsh, and the MAST~Team.
\newblock {Particle confinement of pellet-fuelled tokamak plasma}.
\newblock {\em Nuclear Fusion}, 48(7):075006, 2008.

\bibitem{lang2012aug}
P~T Lang, W~Suttrop, E~Belonohy, M~Bernert, R~M~Mc Dermott, R~Fischer,
  J~Hobirk, O~J W~F Kardaun, G~Kocsis, B~Kurzan, M~Maraschek, P~de~Marne,
  A~Mlynek, P~A Schneider, J~Schweinzer, J~Stober, T~Szepesi, K~Thomsen,
  W~Treutterer, E~Wolfrum, and {the ASDEX Upgrade Team}.
\newblock {High-density H-mode operation by pellet injection and ELM mitigation
  with the new active in-vessel saddle coils in ASDEX Upgrade}.
\newblock {\em Nuclear Fusion}, 52(2):023017, 2012.

\bibitem{baylor2007iter}
L~R Baylor, P~B Parks, T~C Jernigan, J~B Caughman, S~K Combs, C~R Foust, W~A
  Houlberg, S~Maruyama, and D~A Rasmussen.
\newblock {Pellet fuelling and control of burning plasmas in ITER}.
\newblock {\em Nuclear Fusion}, 47(5):443, 2007.

\bibitem{polevoi2005iter}
A~R Polevoi, M~Shimada, M~Sugihara, Yu~L Igitkhanov, V~S Mukhovatov, A~S
  Kukushkin, S~Yu Medvedev, A~V Zvonkov, and A~A Ivanov.
\newblock {Requirements for pellet injection in ITER scenarios with enhanced
  particle confinement}.
\newblock {\em Nuclear Fusion}, 45(11):1451, 2005.

\bibitem{yamada2000pellets}
H~Yamada, R~Sakamoto, Y~Oda, T~Hiramatsu, M~Kinoshita, M~Ogino, R~Matsuda,
  S~Sudo, S~Kato, P~W Fisher, L~R Baylor, and M~Gouge.
\newblock {Development of pellet injector system for large helical device}.
\newblock {\em Fusion Engineering and Design}, 49–50:915, 2000.

\bibitem{sakamoto2006repet}
R~Sakamoto, H~Yamada, Y~Takeiri, K~Narihara, T~Tokuzawa, H~Suzuki, S~Masuzaki,
  S~Sakakibara, S~Morita, M~Goto, B~J Peterson, K~Matsuoka, N~Ohyabu, A~Komori,
  O~Motojima, and {the LHD experimental group}.
\newblock {Repetitive pellet fuelling for high-density/steady-state operation
  on LHD}.
\newblock {\em Nuclear Fusion}, 46(11):884, 2006.

\bibitem{mccarthy2008pinjector}
K~J McCarthy, S~K Combs~S K, L~R Baylor, J~B~O Caughman, D~T Fehling, C~R
  Foust, J~M McGill, J~M Carmona, and D~A Rasmussen.
\newblock {A compact flexible pellet injector for the TJ-II stellarator}.
\newblock {\em Review of Scientific Instruments}, 79(10), 2008.

\bibitem{combs2012pinjector}
S~K Combs, C~R Foust, J~M McGill, J~B~O Caughman, K~J McCarthy, L~R Baylor~M
  Chamorro, D~T Fehling, R~Garc\'ia, J~H Harris, J~Hern\'andez-S\'anchez,
  C~Hidalgo, S~J Meitner, D~A Rasmussen, and R~Unamuno.
\newblock {Results from Laboratory Testing of a New Four-Barrel Pellet Injector
  for the TJ-II Stellarator}.
\newblock {\em Fusion Science and Technology}, 64(3), 2012.

\bibitem{sunnpedersen2015op11}
T~Sunn Pedersen, T~Andreeva, H-S Bosch, S~Bozhenkov, F~Effenberg, M~Endler,
  Y~Feng, D~A Gates, J~Geiger, D~Hartmann, H~H\"olbe, M~Jakubowski, R~K\"onig,
  HP~Laqua, S~Lazerson, M~Otte, M~Preynas, O~Schmitz, T~Stange, Y~Turkin, and
  {the W7-X Team}.
\newblock {Plans for the first plasma operation of Wendelstein 7-X}.
\newblock {\em Nuclear Fusion}, 55(12):126001, 2015.

\bibitem{mccarthy2015ishw}
K~J McCarthy, J~Baldzuhn, R~Sakamoto, A~Dinklage, S~Cats, G~Motojima,
  N~Panadero, B~P\'egouri\'e, H~Yamada, E~Ascas\'ibar, {the LHD Team}, {the
  W7-X Team}, and {the TJ-II Team}.
\newblock {Comparative Study of Pellet Fuelling in 3-D Magnetically Confined
  Plasma Devices}.
\newblock In {\em 20th International Stellarator/Heliotron Workshop,
  Greifswald, Germany}, 2015.

\bibitem{panadero2016phd} 
  N~Panadero, K~J~McCarthy, E~de la Cal, J~Hern\'andez-S\'anchez,
  R~Garc\'ia, M~Navarro and {the TJ-II team}.  
\newblock {Observation of Cryogenic Hydrogen Pellet with a Fast-frame
  Camera System in the TJ-II stellarator}.   
\newblock In {\em 43rd European Physical Society Conference on Plasma Physics, Leuven, Belgium}, 2016.

\bibitem{parks1977ngs}
P~B Parks, R~J Turnbull, and C~A Foster.
\newblock {A model for the ablation rate of a solid hydrogen pellet in a
  plasma}.
\newblock {\em Nuclear Fusion}, 17(3):539, 1977.

\bibitem{sakamoto2001ngs} 
  R~Sakamoto, H~Yamada, K~Tanaka, K~Narihara, S~Morita,
  S~Sakakibara, S~Masuzaki, S~Inagaki, L~R~Baylor, P~W~Fisher,
  S~K~Combs, M~J~Gouge, S~Kato, A~Komori, O~Kaneko, N~Ashikawa,
  P~de Vrxxies, M~Emoto, H~Funaba, M~Goto, K~Ida, H~Idei,
  K~Ikeda, M~Isobe, S~Kado, K~Kawahata, K~Khlopenkov, S~Kubo,
  R~Kumazawa, T~Minami, J~Miyazawa, T~Morisaki, S~Murakami,
  S~Muto, T~Mutoh, Y~Nagayama, Y~Nakamura, H~Nakanishi,
  K~Nishimura, N~Noda, T~Notake, T~Kobuchi, Y~Liang, S~Ohdachi,
  N~Ohyabu, Y~Oka, M~Osakabe, T~Ozaki, R~O~Pavlichenko,
  B~J~Peterson, A~Sagara, K~Saito, H~Sasao, M~Sasao, K~Sato,
  M~Sato, T~Seki, T~Shimozuma, M~Shoji, S~Sudo, H~Suzuki,
  M~Takechi, Y~Takeiri, N~Tamura, K~Toi, T~Tokuzawa, Y~Torii,
  K~Tsumori, I~Yamada, S~Yamaguchi, S~Yamamoto, Y~Yoshimura,
  K~Y~Watanabe, T~Watari, K~Yamazaki, Y~Hamada, O~Motojima and
  M~Fujiwara.  \newblock
  \newblock{Impact of pellet injection on extension of the operational
    region in LHD}
 \newblock {\em Nuclear Fusion}, 41(4):381, 2001.

\bibitem{feng2014emc3}
Y~Feng, H~Frerichs, M~Kobayashi, A~Bader, F~Effenberg, D~Harting, H~Hoelbe,
  J~Huang, G~Kawamura, J~D Lore, T~Lunt, D~Reiter, O~Schmitz, and D~Sharma.
\newblock {Recent Improvements in the EMC3-Eirene Code}.
\newblock {\em Contributions to Plasma Physics}, 54(4-6):426, 2014.

\bibitem{satake2014eps}
S~Satake, {J L Velasco}, A~Dinklage, M~Yokoyama, Y~Suzuki, C~D Beidler,
  H~Maassberg, J~Geiger, A~Wakasa, S~Matsuoka, S~Murakami, D~L\'opez-Bruna,
  N~Pablant, {LHD Exp Group}, {TJ-II Team}, and {W7-AS Team}.
\newblock {Benchmark of local and non-local neoclassical transport calculations
  in helical configurations}.
\newblock In {\em 41th European Physical Society Conference on Plasma Physics,
  Berlin, Germany}, 2014.

\bibitem{velasco2014eps}
{J L Velasco}, D~L\'opez-Bruna, S~Satake, E~Ascas\'ibar, A~Dinklage, T~Estrada,
  K~J McCarthy, F~Medina, M~Ochando, and M~Yokoyama.
\newblock {Validation of local and non-local neoclassical predictions for the
  radial transport of plasmas of low ion collisionallity}.
\newblock In {\em 41th European Physical Society Conference on Plasma Physics,
  Berlin, Germany}, 2014.

\bibitem{hirshman1986dkes}
S~P Hirshman, K~C Shaing, W~I van Rij, C~O Beasley, and E~C Crume.
\newblock {Plasma transport coefficients for nonsymmetric toroidal confinement
  systems}.
\newblock {\em Physics of Fluids}, 29(9):2951--2959, 1986.

\bibitem{satake2006fortec3d}
S~Satake, M~Okamoto an~N~Nakajima, H~Sugama, and M~Yokoyama.
\newblock {Non-Local Simulation of the Formation of Neoclassical Ambipolar
  Electric Field in Non-Axisymmetric Configurations}.
\newblock {\em Plasma and Fusion Research}, 1:002, 2006.

\bibitem{satake2015ishw}
S~Satake, J~L Velasco, D~L\'opez-Bruna, A~Dinklage, J~M Garc\'ia-Rega{\~n}a,
  C~D Beidler, H~Maassberg, J~Geiger, M~Yokoyama, T~Ido, A~Shimizu, K~Tanaka,
  Y~Suzuki, A~Ishizawa, S~Matsuoka, A~Wakasa, S~Murakami, and N~Pablant.
\newblock {Validation and verification of neoclassical transport codes for
  heliotron/stellarator devices}.
\newblock In {\em 20th International Stellarator-Heliotron Workshop,
  Greifswald, Germany}, 2015.

\bibitem{sanchez2013tj-ii}
J~S{\'a}nchez and {TJ-II Team}.
\newblock {Dynamics of flows and confinement in the TJ-II stellarator}.
\newblock {\em Nuclear Fusion}, 53(10):104016, 2013.

\bibitem{mccarthy2015epsd} K~J~McCarthy, N~Panadero, J~L~Velasco,
  S~Combs, J~Caughman, E~de~la~Cal, D~Fehling, C~Foust, R~Garc\'ia,
  J~Hern\'andez-S\'achez, F~Mart\'in, J~McGill, M~Navarro,
  I~Pastor~and~M~C~Rodr\'iguez.  
  \newblock{A pellet injector and associated plasma diagnostics for
    performing plasma studies and fuelling in the TJ-II stellarator}.
 \newblock In {\em 1st EPS Conference on Plasma
    Diagnostics, Frascati, Italy}, 2015.

\bibitem{mccarthy2015cwgm} K~J~McCarty, N~Panadero, J~L~Velasco,
  J~Hern\'andez, E~S\'anchez, R~Garc\'ia~, I~Pastor, E~{de~la~Cal},
  J~M~Fontdecaba, {TJ-II~Team} and {HIBP~Team}.  
  \newblock{A Pellet Injector For Performing Plasma Fueling Studies in
    TJ-II}.
  \newblock In{\em 14th Coordinated Working Group Meeting, Warsaw,
    Poland}, 2015.
  \url{http://ishcdb.nifs.ac.jp/CWGM14/Friday19/F2-14thCGWM_KJM.pdf}

\bibitem{herranz2003TS}
J~Herranz, F~Castej\'on, I~Pastor, and K~J McCarthy.
\newblock {The spectrometer of the high-resolution multiposition Thomson
  scattering diagnostic for TJ-II}.
\newblock {\em Fusion Engineering and design}, 65(4):525--536, 2003.

\bibitem{sanchez2004interferometer}
M~S\'anchez, J~S\'anchez, T~Estrada, E~S\'anchez, P~Acedo, and H~Lamela.
\newblock {High resolution CO2 interferometry on the TJ-II stellarator by using
  an ADC-based phase meter}.
\newblock {\em Review of Scientific Instruments}, 75(10):3414--3416, 2004.

\bibitem{baiao2010SXR}
D~Baiao, F~Medina, M~Ochando, I~Pastor, C~Varandas, A~Molinero, and
  J~Ch\'ercoles.
\newblock {Implementation of multifilter based twin-prototypes for core
  electron temperature measurements in the TJ-II stellarator}.
\newblock {\em Review of Scientific Instruments}, 81(10):10D711, 2010.

\bibitem{baiao2012SXR} D~Baiao, F~Medina, M~Ochando, I~Pastor,
  C~Varandas, A~Molinero, and J~Ch\'ercoles.  
  \newblock {Central electron temperature estimations of TJ-II neutral
    beam injection heated plasmas based on the soft x ray multi-foil
    technique}.  
  \newblock {\em Review of Scientific Instruments}, 83(5):053501, 2012.

\bibitem{fontdecaba2014NPA}
J~M Fontdecaba, S~Ya Petrov, V~G Nesenevich, A~Ros, F~V Chernyshev, K~J
  McCarthy, and J~M Barcala.
\newblock {Upgrade of the neutral particle analyzers for the TJ-II
  stellarator}.
\newblock {\em Review of Scientific Instruments}, 85(11):11E803, 2014.

\bibitem{velasco2011bootstrap}
J~L Velasco, K~Allmaier, A~L\'opez Fraguas, C~D Beidler, H~Maa{\ss}berg,
  W~Kernbichler, F~Castej\'on, and J~A Jim\'enez.
\newblock {Calculation of the bootstrap current profile for the TJ-II
  stellarator}.
\newblock {\em Plasma Physics and Controlled Fusion}, 53(11):115014, 2011.

\bibitem{teubel1994fafner}
A~Teubel and F~P Penningsfeld.
\newblock {Influence of radial electric fields on the heating efficiency of
  neutral beam injection in the W7-AS stellarator}.
\newblock {\em Plasma Physics and Controlled Fusion}, 36(1):143, 1994.

\bibitem{reiter2001eirene}
D~Reiter, M~Baelmans, and P~Boerner.
\newblock {The EIRENE and B2-EIRENE Codes}.
\newblock {\em Fusion Science and Technology}, 47:172, 2001.

\bibitem{fontdecaba2010pfr}
J~M Fontdecaba, I~Pastor, J~Ar\'evalo, J~Herranz, K~J McCarthy, and
  G~S\'anchez-Burillo.
\newblock {Comparisons of Electron Temperature and Density, and Ion Temperature
  Profiles in the TJ-II Stellarator}.
\newblock {\em Plasma Fusion Research}, 5:S2085, 2010.

\bibitem{milligen2011bayes}
B~Ph van Milligen, T~Estrada, E~Ascas\'ibar, D~Tafalla, D~L\'opez-Bruna,
  A~L\'opez Fraguas, J~A. Jim\'enez, I~Garc\'ia-Cort\'es, A~Dinklage, and
  R~Fischer.
\newblock {Integrated data analysis at TJ-II: The density profile}.
\newblock {\em Review of Scientific Instruments}, 82(7):073503, 2011.

\bibitem{pereverzev2002ASTRA}
G~V Pereverzev and P~Yushmanov.
\newblock {ASTRA automated system for transport analysis}.
\newblock {\em IPP Technical Report}, 5/98, 2002.

\bibitem{satake2016ppcf}
S~Satake, {J L Velasco}, D~L\'opez-Bruna, A~Dinklage, J~M Garc\'ia-Rega{\~n}a,
  C~D Beidler, H~Maassberg, J~Geiger, M~Yokoyama, T~Ido, A~Shimizu, K~Tanaka,
  Y~Suzuki, A~Ishizawa, S~Matsuoka, A~Wakasa, S~Murakami, and N~Pablant.
\newblock {Validation and verification of neoclassical transport codes for
  heliotron/stellarator devices}.
\newblock {\em Plasma Physics and Controlled Fusion}, 2016, submitted.

\bibitem{velasco2012prl}
J~L Velasco, J~A Alonso, I~Calvo, and J~Ar\'evalo.
\newblock {Vanishing neoclassical viscosity and physics of the shear layer in
  stellarators}.
\newblock {\em Physical Review Letters}, 109:135003, 2012.

\end{thebibliography}

%\section*{References}
%\bibliographystyle{unsrt}

 \end{document}